\documentstyle[11pt,fleqn]{article}

\begin{document}

\title{Pad\'e-Improvement of QCD Running Coupling Constants, Running Masses, 
Higgs Decay Rates, and Scalar Channel Sum Rules}

\author{V. Elias\thanks{Permanent Address: Department of
Applied Mathematics, The University of Western Ontario, London,
Ontario N6A 5B7, CANADA} and T. G. Steele\\
Department of Physics and Engineering Physics\\
University of Saskatchewan\\
Saskatoon, Saskatchewan S7N 5C6\\
Canada\\
and\\
F. Chishtie, R. Migneron and K. Sprague\\
Department of Applied Mathematics\\
The University of Western Ontario\\
London, Ontario N6A 5B7\\
Canada
}
\maketitle

\smallskip


\begin{abstract}We discuss Pad\'e-improvement of known four-loop order results based
upon an asymptotic three-parameter error formula for
Pad\'e-approximants. We derive an explicit formula estimating the
next-order coefficient $R_4$ from the previous coefficients in a
series $1+R_1 x + R_2x^2 + R_3x^3$. 
We show that such an estimate is within 0.18\%
of the known five-loop order term in the $O(1)$ $\beta$-function, and
within 10\% of the known five-loop term in the $O(1)$ anomalous
mass-dimension function $\gamma_m(g)$. We apply the same formula to
generate a [2$\vert$2] Pad\'e-summation of the QCD $\beta$-function 
and anomalous mass dimension in
order to demonstrate both the relative insensitivity of the evolution of
$\alpha_s(\mu)$ and the running quark masses to higher order corrections, 
as well as a somewhat increased
compatibility of the present empirical range for $\alpha_s(m_\tau)$
with the range anticipated via evolution from the present empirical
range for $\alpha_s(M_z)$.  
For $3 \leq n_f \leq 6$ we demonstrate that positive zeros of
any [2$\vert$2] Pad\'e-summation estimate of the
all-orders $\beta$-function which incorporates known
two-, three-, and four-loop contributions necessarily 
correspond to {\it ultraviolet} fixed points, {\it regardless} 
of the unknown five-loop term.
Pad\'e-improvement of higher-order perturbative expressions 
is presented for the
decay rates of the Higgs into two gluons and into a $b \bar{b}$ pair,
and is used to show the relative insensitivity of these rates to
higher order effects.  However, Pad\'e-improvement
of the purely-perturbative component of scalar/pseudoscalar current 
correlation functions is indicative of 
large theoretical uncertainties in QCD sum rules for these
channels, particularly if the continuum-threshold parameter $s_0$ is near
1~GeV$^2$.
\end{abstract}
\newpage

\section*{1. Introduction}

A recent body of work [1-3] has demonstrated how higher order terms in a 
number of field-theoretical perturbative series can be estimated by 
Pad\'e-approximant techniques.

Of particular
interest are applications to QCD quantities, particularly $\beta$- and 
$\gamma$-functions now known to four-
loop order in $\alpha_s$ [4-6].  Pad\'e-approximant methods have already addressed the 
$n_f$ - (flavour-number-) dependence of higher order terms in the QCD
$\beta$- and $\gamma$-functions [2,3].  Near cancellations
between coefficients of successive powers of $n_f$, however, can lead to large uncertainty in the
estimated overall size of such higher-loop contributions -- small errors in fitted coefficients of
$n_f^k$ have been seen to lead to much larger errors in the aggregate (now known) four-loop
contribution to the $\beta$-function [2].

      In the present paper, our focus will be on using Pad\'e-approximant methods to estimate
the magnitude of higher-order corrections to quantities already calculated to three- and 
four-loop order in QCD.  We assess the theoretical uncertainty of such calculations by seeing
how
closely they coincide with their own Pad\'e-improvements, as well as whether successive orders
of perturbation theory exhibit convergence toward Pad\'e-summation estimates of the full
perturbative series.

      The present paper is phenomenologically oriented, specifically aimed at developing
Pad\'e-improved estimates of what are hoped to be computationally
and/or experimentally accessible quantities.  The
particular items of interest considered are the $O(N)$
$\beta$-function and anomalous mass dimension (Section 2),
the running QCD coupling constant (Section 3), the running
quark masses (Section 4), the Higgs-boson decay rates into two gluons and into a
$b\bar{b}$ pair
(Section 5), and the purely perturbative content of QCD-sum rules
based upon scalar
and pseudoscalar current correlation functions (Section 6). In Section 3, we also discuss some
general implications of Pad\'e-summation as an approximation to all orders of perturbation
theory. 
In particular, we analyze the fixed point structure of the most general
[2 $\vert$ 2] Pad\'e-summation
estimates of the full content of QCD $\beta$-functions for $n_f =
\{3,4,5,6\}$ whose Maclaurin expansions
coincide with presently known perturbative contributions.

      In the section that immediately follows, we discuss how
Pad\'e-approximant methods can be
used to estimate higher-order corrections to perturbative series in which the
leading four terms are known.  Although this methodology appears in other work [3], the
presentation leading to eq. (2.12) [which has not appeared elsewhere] will, 
it is hoped, be of some value to those unfamiliar with Pad\'e-approximant 
methods.  We also demonstrate the reasonable agreement of predictions
based on (2.12) with now-known five-loop terms in $O(N)$ $\beta$- and
$\gamma$-functions.  

\section*{2. The APAP Algorithm: Pad\'e-Improvement for Pedestrians}

\renewcommand{\theequation}{2.\arabic{equation}}
\setcounter{equation}{0}

We consider the general problem of developing a Pad\'e-improvement of the series
\begin{equation}
S \equiv 1 + R_1 x + R_2 x^2 + R_3 x^3 + ... ,
\end{equation}
where $\{R_1, R_2, R_3 \}$ are known and $\{ R_4, R_5, ...\}$ are not
known, through use of the asymptotic error formula for estimating
$R_{N+M+1} \; x^{N+M+1}$ via the Pad\'e approximant
\begin{eqnarray}
S_{[N|M]} & \equiv & \frac{1+a_1 x + a_2 x^2 + ... a_N x^N}{1 + b_1 x +
b_2 x^2 + ... + b_M x^M} \nonumber \\
& = & 1 + R_1 x + R_2 x^2 + R_3 x^3 + ... + R_{N+M+1} \; x^{N+M+1} +
...
\end{eqnarray}
Let $R_{N+M+1}^{Pad\acute{e}}$ be the prediction one would obtain from the
$[N|M]$ Pad\'e Approximant, and let $R_{N+M+1}$ be the true value of
the coefficient.  The structure of the asymptotic error formula is
given by [2]
\begin{equation}
\frac{R_{N+M+1}^{Pad\acute{e}} - R_{N+M+1}} {R_{N+M+1}} = - \frac{M!
A^M}{[N+M+aM+b]^M}
\end{equation}
with numbers $\{A,a,b\}$ (independent of $N,M$) to be determined.

This error formula simplifies considerably if $M$ is always chosen to
be 1 [3]:  the right hand side of (C) becomes $-A / [N+1+(a+b)]$, with
only the two numbers $\{A, a+b\}$ to be determined.  These can be
determined explicitly for the series $S$ in (2.1) given knowledge of the three
coefficients $\{R_1, R_2, R_3 \}$.

Given knowledge of $R_1$ only, the $[0|1]$ Pad\'e approximant
\begin{equation}
S_{[0|1]} = \frac{1}{1+b_1 x} = 1 - b_1 x + b_1^2 x^2 ... = 1 + R_1 x
+ R_2^{Pad\acute{e}} x^2
\end{equation}
predicts $R_2^{Pad\acute{e}} = R_1^2$.  Consequently, we see from the
asymptotic error formula (2.3) that
\begin{equation}
\frac{R_2^{Pad\acute{e}} - R_2}{R_2} = \frac{R_1^2 - R_2}{R_2} = \frac{
-A}{1+(a+b)} \equiv \delta_2
\end{equation}
Given knowledge of only $R_1$ and $R_2$, the $[1|1]$ Pad\'e
approximant
\begin{eqnarray}
S_{[1|1]} = \frac{1 + a_1 x}{1 + b_1 x} & = & 1 + (a_1 - b_1) x +
b_1(b_1 - a_1) x^2 + b_1^2 (a_1 - b_1) x^3 + ... \nonumber \\
& = & 1 + R_1 x + R_2 x^2 + R_3^{Pad\acute{e}} x^3 ...
\end{eqnarray}
predicts $R_3^{Pad\acute{e}} = R_2^2 / R_1$.  The asymptotic error
formula (2.3) for this case implies that
\begin{equation}
\frac{R_2^2 / R_1 - R_3}{R_3} = \frac{-A}{2 + (a+b)} \equiv \delta_3
\end{equation}

Given knowledge of $\{R_1, R_2, R_3\}$ in the series (2.1), the
relative errors $\delta_2$ and $\delta_3$ are specified completely
by the left-hand sides of (2.5) and (2.7).  These two equations may
be regarded as two equations in the two unknowns $A$ and $(a+b)$
characterizing the asymptotic error formula (2.3) when $M=1$.  The
solution to these two equations is
\begin{equation}
A = \left[ 1 / \delta_2 - 1 / \delta_3 \right]^{-1},
\end{equation}

\begin{equation}
(a+b) = \frac{\delta_2 - 2 \delta_3}{\delta_3 - \delta_2} \; .
\end{equation}
This information is sufficient to generate an asymptotic Pad\'e
approximant (APAP)
estimate [2,3] of the unknown coefficient $R_4$ in the series (2.1). 
Consider the $[2|1]$ Pad\'e approximant
\begin{eqnarray}
S_{[2|1]} & = & \frac{1+a_1 x + a_2 x^2}{1+b_1 x} \nonumber \\
& = & 1 + (a_1 - b_1) x + [a_2 - b_1 (a_1 - b_1)] x^2 \nonumber \\
&+ & [-b_1 [ a_2 - b_1 (a_1 - b_1) ] ] x^3 \nonumber \\
& + & [b_1^2 [a_2 - b_1 ( a_1 - b_1 )] ]x^4 + ... \nonumber \\
& = & 1 + R_1 x + R_2 x^2 + R_3 x^3 + R_4^{Pad\acute{e}} x^4
\end{eqnarray}
The three known values of $\{R_1, R_2, R_3\}$ completely determine
the three parameters $\{ a_1, a_2, b_1\}$ characterizing $S_{[2|1]}$. 
We see from (2.10) that $R_4^{Pad\acute{e}} = R_3^2 / R_2$.  However,
the asymptotic error formula (2.3) suggests that a more accurate
estimate of the true value $R_4$ differs from $R_4^{Pad\acute{e}}$ by
a predictable relative error:
\begin{equation}
\frac{R_3^2 / R_2 - R_4}{R_4} \equiv \delta_4 = \frac{-A}{3+(a+b)} ,
\end{equation}
in which case we find from (2.11), (2.8) and (2.9) that
\begin{eqnarray}
R_4 & = & \frac{R_3^2 / R_2}{1+ \delta_4} \nonumber \\
& = & \frac{R_3^2 (\delta_2 - 2 \delta_1)}{R_2 (\delta_2 - 2\delta_1
- \delta_1 \delta_2)} \nonumber \\
& = & \frac{R_3^2 (R_2^3 + R_1 R_2 R_3 - 2 R_1^3 R_3)}{R_2 (2 R_2^3 -
R_1^3 R_3 - R_1^2 R_2^2)} .
\end{eqnarray}

As an example, we test the applicability of (2.12) by comparing its
prediction to the known ${\cal O}(g^6)$ coefficient of
the $O(1)$ $\beta$-function [7]:

\begin{eqnarray}
\beta^{(1)}(g) & = & 1.5 g^2 \left[ 1 - (17/9) g + 10.8499 g^2 
\right. \nonumber \\
& - & \left. 90.5353 g^3 + 949.523 g^4 + {\cal O}(g^5) \right] .
\end{eqnarray}
Identifying $R_1 = -17/9, \; R_2 = 10.8499$ and $R_3 = -90.5353$, we
find from (2.12) that $R_4 = 947.8$ in startlingly close agreement to
the next $(g^4)$ term within (2.13).  Although APAP improvement has also
been applied elsewhere [2] to ${\cal O}(g^6)$ terms in $O(N)$
$\beta$-functions, the result obtained here relies on direct and
explicit use of the full asymptotic error formula (2.3).
A comparison of (2.12) predictions and exact values of $\beta_4$, the
$O(N)$ $\beta$-function coefficient of $g^6$, is presented in Table
I.  We emphasize that these predictions are {\it not} obtained by a
fitting of the $N$-dependence or any knowledge of the
$N^4$-dependence of $\beta^{(4)}$, as is the case in Table 3 of ref.
[2];  for
comparative purposes, the predictions of ref. [2] are also listed in
Table I.

We can use (2.12) to predict the known $R_4$
coefficient within the $O(1)$ theory's anomalous mass dimension [7],
as well:

\begin{eqnarray}
\gamma_m (g) & = & (g/2) \left[ 1 - 0.8333 g + 3.500 g^2 \right. \nonumber
\\
& - & \left. 19.96 g^3 + 150.8 g^4 + {\cal O}(g^5) \right]
\end{eqnarray} 
Eq. (2.12) predicts $R_4 = 135.1$, a result only 10\% off the 
150.8 value given in (2.14).  Table I shows that predicted $R_4$ 
coefficients for
$\gamma_m(g)$ within $O(2)$, $O(3)$, and $O(4)$ also remain within
20\% of their true values, as given in [7]. These results provide a
reasonable basis for applying (2.12) [and its concomitant asymptotic error
formula (2.3)] to the $\beta$- and $\gamma$-functions of QCD, as we
will do in Sections 3 and 4.

A final improvement of the series (2.1) is possible by expressing this series as
a $[2|2]$ diagonal approximant --- this is a more accurate
representation of the infinite series $S$ than one would obtain by
arbitrarily truncating the series after the $R_4 x^4$ term.  Given
known values of $\{R_1, R_2, R_3\}$ and using the APAP estimate
(2.12) for $R_4$, the approximant $S_{[2|2]}$ of the infinite series
$S$ is fully determined:
\begin{equation}
S \rightarrow S_{[2|2]} = \frac{1+a_1 x + a_2 x^2}{1 + b_1x + b_2x^2}
\; ,
\end{equation}
\begin{equation}
b_1 = \frac{R_1 R_4 - R_2 R_3}{R_2^2 - R_1 R_3} \; ,
\end{equation}
\begin{equation}
b_2 = \frac{R_3^2 - R_2 R_4}{R_2^2 - R_1 R_3} \; ,
\end{equation}
\begin{equation}
a_1 = R_1 + b_1 \; ,
\end{equation}
\begin{equation}
a_2 = R_2 + b_1 R_1 + b_2 \; .
\end{equation}
Eq. (2.15), as determined from (2.12) and (2.16-19), constitutes the
procedure we denote as ``Pad\'e-summation'' of the series $S$ in
(2.1).

\renewcommand{\theequation}{3.\arabic{equation}}
\setcounter{equation}{0}

\section*{3.  Pade-Improvement of the QCD Coupling}

      The QCD minimal subtraction (MS or $\overline{MS}$) renormalization-group functions
$\beta(x)$ and
$\gamma (x)$ are now known to 4-loop
order [4,5,6].  Prior work involving Pad\'e-improvement methods has attempted to predict the
flavour-dependence of these functions to 4- and 5- loop order [2,3].
$\beta_3$  and $\beta_4$, the 4- and 5-
loop order corrections to the $\beta$-function, are respectively third- and fourth-degree
polynomials in
$n_f$, and Pad\'e methods have already shown some success in predicting the polynomial
coefficients
now known for $\beta_3$  [2].  However, an accurate determination of the polynomial
coefficients within
$\beta_3$ is not reflected in the accuracy with which $\beta_3$ is itself determined.  
Thus the overall Pad\'e-
driven estimate of $\beta_3$ presented in [2] for $n_f  = 3 \; [ \beta_3
= (7.6 \pm 0.1)
\cdot 10^3 /256 = 30 \pm 1$, using
normalization conventions appropriate for (3.1) below] is substantially below the true value
($\beta_3 = 47.23$), even after allowances are made for claimed uncertainties arising from
quadratic Casimir contributions [3].  It should also be noted that this estimate, once disentangled
from a fitting procedure aimed at ascertaining the explicit $n_f$ dependence of
$\beta_3$, follows from a simplified version of the asymptotic error formula 
(2.3), in which the denominator $(N + M + aM + b)^M$  is taken to be $N^M$
[2].
Indeed, it is difficult to tell at this stage whether the discrepancy
between Pad\'e-estimates of $\beta_3$ and the true value arises primarily 
from quadratic-Casimir contributions not occurring in lower orders, or 
alternatively, arises from the error involved in simplifying the asymptotic 
error formula in order to make an estimate of $\beta_3$  possible.  Without
such a simplification, the error formula (2.3) has (in principle) three arbitrary constants (A,a,b)
instead of one
(A).

Consequently, in this section we will predict $\beta_4$ directly using the 
APAP-algorithm (2.12)
following from the full asymptotic error formula (2.3).  We will not prejudice these predictions
of $\beta_4$  by attempting a fit of the polynomial dependence on
$n_f$, nor will we attempt to disentangle quadratic-and-higher Casimir 
contributions from $\beta_3$ and $\beta_4$.  Rather, we will make
distinct predictions of $\beta_4$ via (2.12) for $n_f = {3,4,5,6}$.  The validity of such an
approach,
particularly the possibility that the full asymptotic error formula is inclusive of higher-order
Casimir contributions to $\beta(x)$, would be best established by comparison to an exact
calculation
of $\beta_4$, when available.

      We define the $\beta$-function as in [4]:

\renewcommand{\theequation}{3.1\alph{equation}}
\setcounter{equation}{0}

\begin{eqnarray}
\mu^2 \frac{d}{d \mu^2} x & = & - x^2 \sum_{i=0}^\infty \beta_i x^i
\nonumber \\
& = & - \beta_0 x^2 \sum_{i = 0}^\infty R_i x^i,
\end{eqnarray}

\noindent with $x \equiv  \alpha_s(\mu)/\pi$ and

\begin{equation}
R_i \equiv \beta_i / \beta_0,
\end{equation}
Known values of $\beta_0 - \beta_3$ are as follows [4,5]:

\renewcommand{\theequation}{3.2\alph{equation}}
\setcounter{equation}{0}

\begin{equation}
\beta_0 = (11 - 2n_f / 3) / 4,
\end{equation}
\begin{equation}
\beta_1 = (102 - 38 n_f / 3) / 16,
\end{equation}
\begin{equation}
\beta_2 = (2857/2 - 5033 n_f / 18 + 325 n_f^2 / 54) / 64,
\end{equation}
\begin{eqnarray}
\beta_3 & = & 114.23033 - 27.133944 n_f + 1.5823791 n_f^2 \nonumber
\\
& + & 5.85669582 \cdot 10^{-3} n_f^3.
\end{eqnarray}

We use (3.2), (3.1b), and (2.12) to predict the following values for
$\beta_4$:

\renewcommand{\theequation}{3.\arabic{equation}}
\setcounter{equation}{2}

\begin{equation}
n_f = 3: \; \; \; R_4 = -849.74, \; \; \beta_4 = -1911.9;
\end{equation}
\begin{equation}
n_f = 4: \; \; \; R_4 = 40.203, \; \; \beta_4 = 83.7563;
\end{equation}
\begin{equation}
n_f = 5: \; \; \; R_4 = 70.203, \; \; \beta_4 = 134.56;
\end{equation}
\begin{equation}
n_f = 6: \; \; \; R_4 = -239.22, \; \; \beta_4 = -418.64.
\end{equation}

The large negative values for $n_f=3$  and $n_f=6$ reflect the near cancellation 
of the factor $(2R_2^3 - R_1^3 R_3 - R_1^2 R_2^2)$ in the denominator of (2.12).  
Since a change of sign can easily occur if this cancellation is over- or 
under-estimated, the safest interpretation of (3.3) and (3.6) is to predict
a relatively large magnitude for $\beta_4$, with the sign uncertain.  

Corresponding results from Table III of reference [3] with quadratic
Casimir contributions and
a
1/1024 normalization factor appropriate to (3.1a) included are
$\beta_4 = \{278 (n_f=3), 202 (n_f=4),165 (n_f=5), 166 (n_f=6)\}$.  As stated 
earlier, these latter results are based upon a fit to the
polynomial coefficients of $n_f^k$ in $\beta_4$, with near cancellation of very large
opposite-sign
coefficients for the $k = 0$ and $k = 1$ terms.  These latter results appear most 
consistent with those obtained above when $n_f=5$.

To get a feeling of the magnitude of these Pad\'e estimates of 5-loop effects, we generate
the [2$\vert$2] approximant (2.15) from the known values for $R_1, R_2,$ and
$R_3$ and our estimate of $R_4$, and incorporate this approximant directly 
into the $\beta$-function $[\mu^2 dx/d\mu^2 \equiv \beta(x)]$:

\renewcommand{\theequation}{3.7\alph{equation}}
\setcounter{equation}{0}

\begin{equation}
n_f = 3: \; \; \; \beta(x) = - \frac{9x^2}{4} \left[ \frac{1 +
94.383x - 75.605x^2}{1 + 92.606x - 244.71x^2} \right],
\end{equation}
\begin{equation}
n_f = 4: \; \; \; \beta(x) = - \frac{25x^2}{12} \left[ \frac{1 -
5.8963x - 4.0110x^2}{1 - 7.4363x + 4.3932x^2} \right],
\end{equation}
\begin{equation}
n_f = 5: \; \; \; \beta(x) = - \frac{23x^2}{12} \left[ \frac{1 -
5.9761x - 6.9861x^2}{1 - 7.2369x - 0.66390x^2} \right].
\end{equation}

We can use these Pad\'e approximations to the full $\beta$-function to evolve 
$\alpha_s(\mu) = \pi x(\mu)$ down to $\mu = 1$ GeV from an initial
condition $\alpha_s(M_z)$ = 0.118 [8] through use of the following (4- and 5-) flavour threshold
matching conditions with $m(\mu_{th}) = \mu_{th}$ [9]:

\renewcommand{\theequation}{3.\arabic{equation}}
\setcounter{equation}{7}

\begin{eqnarray}
x^{(n_f - 1)} (\mu_{th}) & = & x^{(n_f)} (\mu_{th}) \left[ 1 + 0.1528 \left(
x^{(n_f)} (\mu_{th}) \right)^2 \right. \nonumber \\
& + & \left. \left\{ 0.9721 - 0.0847 (n_f - 1) \right\} \left(
x^{(n_f)} (\mu_{th}) \right)^3 \right].
\end{eqnarray}

In Figure 1 we display the evolution of $\alpha_s(\mu)$ to low
energies through use of two-loop, three-loop, four-loop, and the
Pad\'e-improved four-loop $\beta$-functions of (3.7).  All of these
curves are generated from the initial condition $\alpha_s(M_z) =
0.118$, with $\mu_{th} = 4.3$ GeV ({\it i.e.} $m_b (m_b) =
4.3$ GeV) identified as the $n_f = 5$ flavour
threshold, and with $\mu_{th} = 1.3$ GeV identified as the $n_f = 4$
flavour threshold.  Eq. (3.8) is utilized in full
in both the four-loop and Pad\'e improved calculations;
it is utilized to ${\cal O}(x^2)$ to generate flavour-threshold initial 
conditions in the three-loop calculation, and the matching condition
is trivial for the two-loop calculation.  
It is evident from the figure that curves
from successive orders of the $\beta$-function appear to converge
from below to that generated via (3.7), the Pad\'e-summation
approximating all orders.  The gaps between curves of successive
order clearly narrow as the order increases.  Figure 1 shows that
the Pad\'e-summation leads to a curve for $\alpha_s(\mu)$ that
exceeds the unimproved four-loop curve by less than 1\%.  Such a
difference is inconsequential compared to the estimated uncertainties
in $\alpha_s(M_z)$ and $\mu_{th}$ for four- and five-flavour thresholds.  
Both the four-loop and the Pad\'e-improved curve are
indicative of benchmark values $\alpha_s(1 GeV) = 0.48$ and
$\alpha_s(m_\tau) = 0.32$.  However, if the flavour thresholds and
$\alpha_s(M_z)$ are assigned their accepted [8] lower-bound values
[$\alpha_s(M_z) = 0.115$, $\mu_{th}^{(n_f = 5)} = 4.1$ GeV,
$\mu_{th}^{(n_f=4)} = 1.0$ GeV], the values we obtain at $\mu = 1$ GeV
and $\mu = m_\tau$ from either four-loop or Pad\'e-summation
$\beta$-functions are $\alpha_s(1 GeV) = 0.41$ and $\alpha_s(m_\tau) =
0.29$.  Corresponding upper-bound values [$\alpha_s(M_z) =
0.121$, $\mu_{th}^{(n_f = 5)} = 4.5$ GeV, $\mu_{th}^{(n_f = 4)} = 1.6$ GeV]
lead to $\alpha_s(1 GeV) = 0.57$ and $\alpha_s(m_\tau) = 0.35$.  In
view of these (much-) larger-than-1\% uncertainties in low-energy
values, the best possible test at present of Pad\'e-improvement
would be nonempirical, {\it i.e.} a comparison to an explicit 5-loop 
{\it calculation} of the $\beta$-function.

We note, however, that higher-order effects do appear to increase the
overlap between the present (somewhat large) empirical range for
$\alpha_s(m_\tau)$ (0.370 $\pm$ 0.033 [8]) and the range predicted
via evolution down from the present empirical range for
$\alpha_s(M_z)$ (0.118 $\pm$ 0.003 [8]).  Taking into account the
present uncertainty in the five-flavour threshold ($\mu_{th} = 4.3
\pm 0.3$ GeV [8]) and incorporating the matching condition (3.8) for
$\alpha_s(\mu)$ below and above $\mu_{th}$, we find the following
{\it predicted} ranges for $\alpha_s(m_\tau)$ for two-loop,
three-loop, four-loop and Pad\'e-summation $\beta$-functions:

\renewcommand{\theequation}{3.9\alph{equation}}
\setcounter{equation}{0}

\noindent 2--Loop:
\begin{equation}
0.2910 \leq \alpha_s (m_\tau) \leq 0.3391,
\end{equation}
\begin{equation}
[\alpha_s(m_\tau)]_{cv} = 0.3137,
\end{equation}

\renewcommand{\theequation}{3.10\alph{equation}}
\setcounter{equation}{0}

\noindent 3--Loop:
\begin{equation}
0.2944\leq \alpha_s (m_\tau) \leq 0.3451,
\end{equation}
\begin{equation}
[\alpha_s(m_\tau)]_{cv} = 0.3182,
\end{equation}

\renewcommand{\theequation}{3.11\alph{equation}}
\setcounter{equation}{0}

\noindent 4--Loop:
\begin{equation}
0.2957\leq \alpha_s(m_\tau) \leq 0.3477,
\end{equation}
\begin{equation}
[\alpha_s(m_\tau)]_{cv} = 0.3200,
\end{equation}

\renewcommand{\theequation}{3.12\alph{equation}}
\setcounter{equation}{0}

Pad\'e-Improved:
\begin{equation}
0.2963\leq \alpha_s(m_\tau) \leq 0.3489,
\end{equation}
\begin{equation}
[\alpha_s(m_\tau)]_{cv} =  0.3208.
\end{equation}
The ranges listed above progressively overlap the low end of the
present experimental range $0.337 \leq \alpha_s(m_\tau) \leq 0.403$.
The central values ($cv$) displayed above are evolved down from
$\alpha_s(M_z)$ = 0.118 with a five-flavour threshold at $\mu_{th}$ =
4.3 GeV.  The lower bounds evolve from $\alpha_s(M_z)$ = 0.115 with
$\mu_{th}$ = 4.0 GeV, and the upper bounds evolve from
$\alpha_s(M_z)$ = 0.121 with $\mu_{th}$ = 4.6 GeV.  

There are also some unexpected theoretical consequences arising from 
estimating the 
summation of the
full $\beta$-function series through use of an appropriately chosen
Pad\'e approximant [1].  The most general [2$\vert$2] approximant must yield 
$1 + R_1 x + R_2 x^2 + R_3 x^3 + R_4 x^4$ as the first five terms
of its Maclaurin expansion.  For optimal generality, we require that 
$R_1, R_2$, and $R_3$ be given by
(3.1b) and (3.2), but allow $R_4$ to be {\it arbitrary}.  We then find that

\renewcommand{\theequation}{3.\arabic{equation}}
\setcounter{equation}{12}

\begin{equation}
\beta(x) = - \beta_0 x^2 \left[ \frac{1+a_1 x + a_2 x^2}{1 + b_1 x
+ b_2 x^2} \right],
\end{equation}

with $\{a_1, a_2, b_1, b_2\}$ linear in $R_4$ as follows:

\bigskip

\renewcommand{\theequation}{3.14\alph{equation}}
\setcounter{equation}{0}

\noindent$n_f = 3$:
\begin{equation}
a_1 = 7.1945 - 0.10261 R_4,
\end{equation}
\begin{equation}
a_2 = -11.329 + 0.075643 R_4,
\end{equation}
\begin{equation}
b_1 = 5.4168 - 0.10261 R_4,
\end{equation}
\begin{equation}
b_2 = -25.430 + 0.25806 R_4;
\end{equation}

\bigskip

\renewcommand{\theequation}{3.15\alph{equation}}
\setcounter{equation}{0}

\noindent$n_f = 4$:
\begin{equation}
a_1 = 4.8401 - 0.11068 R_4,
\end{equation}
\begin{equation}
a_2 = -8.1842 + 0.10836 R_4,
\end{equation}
\begin{equation}
b_1 = 3.3001 - 0.11068 R_4,
\end{equation}
\begin{equation}
b_2 = -16.314 + 0.21904 R_4
\end{equation}

\bigskip

\renewcommand{\theequation}{3.16\alph{equation}}
\setcounter{equation}{0}

\noindent$n_f = 5$:
\begin{equation}
a_1 = 2.6793 - 0.12329 R_4,
\end{equation}
\begin{equation}
a_2 = -6.19671 - 0.011245 R_4,
\end{equation}
\begin{equation}
b_1 = 1.4184 - 0.12329 R_4,
\end{equation}
\begin{equation}
b_2 = -9.4599 + 0.14421 R_4;
\end{equation}

\bigskip

\renewcommand{\theequation}{3.17\alph{equation}}
\setcounter{equation}{0}

\noindent $n_f = 6$:
\begin{equation}
a_1 = 0.61085 - 0.18424 R_4,
\end{equation}
\begin{equation}
a_2 = -6.6275 - 0.24181 R_4,
\end{equation}
\begin{equation}
b_1 = -0.31772 - 0.18424 R_4,
\end{equation}
\begin{equation}
b_2 = -6.0423 - 0.057574 R_4.
\end{equation}

For all $n_f$ values listed above, the first positive zero of $1 +
a_1 x + a_2 x^2$ in (3.13) is found to be above the first positive zero of 
$1 + b_1 x + b_2 x^2$, {\it regardless} of the choice for $R_4$.  Consequently,
the smallest positive zero of $\beta(x)$, {\it if given by (3.13)}, is necessarily 
an ultraviolet fixed point and
{\it not} an infrared fixed point, an inescapable result of the denominator sign
change for $x$ between
0 and the first positive zero of (3.13).  Moreover, for those values of 
$R_4$ for which a second
positive zero of $1 + a_1 x + a_2 x^2$ is possible, we find from
(3.14-17) that a second positive
zero
of $1 + b_1 x + b_2 x^2$  will also occur at some value of $x$ between the 
two positive zeros of $1 + a_1 x + a_2 x^2$. This ensures that {\it
neither} positive zero of $1 + a_1 x + a_2 x^2$ corresponds to an infrared 
fixed point.

In Figure 2, a schematic diagram is presented showing different branches for the evolution 
of $x(\mu)$ anticipated from a $\beta$-function (3.13) with the above-described 
alternation of positive denominator and numerator zeros, with the smallest 
positive zero occurring for the denominator.  Zeros of $1 + b_1 x +
b_2 x^2$  represent values of $x$ for which $x(\mu)$ has infinite slope.  
Zeros of $1 + a_1 x + a_2 x^2$ are fixed-point values of $x$.  As is evident from the figure,
all such fixed points are necessarily
ultraviolet, and the infrared region is {\it inaccessible} for values of
$\mu$ less than those corresponding
to zeros of the denominator.  Such behaviour suggests...

\begin{description}
\item 1)  ...the possible existence of a strong phase of
QCD at short-distances, reflective of a nonzero {\it ultraviolet}
fixed point, and

\item 2) ... the inapplicability of a perturbative theory of quarks and
gluons to the infrared region, specifically the region excluded from
the domain of the first positive branch of $x(\mu)$ [Fig 2].
\end{description}

The former statement above may have ramifications for scenarios of 
dynamical electroweak symmetry breaking that usually involve 
a distinct technicolour group.  The latter statement
parallels old infrared slavery ideas, except that the inapplicability
of perturbation theory to low energies is not seen to follow from the
coupling constant growing  infinite (or nonperturbatively large), as
in infrared slavery, but from an explicit decoupling of the infrared
region from the ultraviolet by virtue of $\beta$-function
singularities alternating with $\beta$-function zeros
(Fig. 2).  We have verified that this alternation occurs in the most
general [2$\vert$2] Pad\'e-summation of the $\beta$-function even 
when $n_f = 0$.

\section*{4.  Pad\'e-Improvement of the Running Mass}

The running mass $m(\mu)$ satisfies the differential equation

\renewcommand{\theequation}{4.\arabic{equation}}
\setcounter{equation}{0} 

\begin{equation}
\frac{dm}{dx} = m \frac{\gamma(x)}{\beta(x)}
\end{equation}
where $x \equiv \alpha_s(\mu)/\pi$, as in the previous section, and
where $\beta(x) \left( \equiv \mu^2 dx/d\mu^2 \right)$ is given by
(3.1a) and (3.2).  The QCD MS anomalous mass dimension function
$\gamma(x)$ has been calculated to four-loop order [4,6]:

\renewcommand{\theequation}{4.2\alph{equation}}
\setcounter{equation}{0}

\begin{equation}
\gamma(x) = -x[1 + \sum_{i=1} \gamma_i x^i]
\end{equation}

\begin{equation}
\gamma_1 = 4.20833 - 0.138889n_f
\end{equation}

\begin{equation}
\gamma_2 = 19.5156 - 2.28412n_f - 0.0270062n_f^2
\end{equation}

\begin{equation}
\gamma_3 = 98.9434 - 19.1075 n_f + 0.276163 n_f^2 - 0.00579322 n_f^3.
\end{equation}
Pad\'e improvement of the square-bracketed expression within (4.2a)
is straightforward via the methods of Section 2 by identifying $R_1,
R_2$ and $R_3$ in (2.1) with $\gamma_1, \gamma_2$ and $\gamma_3$. 
Using (2.12) we obtain the following APAP estimates for $\gamma_4$:

\renewcommand{\theequation}{4.\arabic{equation}}
\setcounter{equation}{2}
 
\begin{equation}
n_f = 3: \; \; \; \gamma_4 = 162.987
\end{equation}

\begin{equation}
n_f = 4: \; \; \; \gamma_4 = 75.2349
\end{equation}

\begin{equation}
n_f = 5: \; \; \; \gamma_4 = 12.5550
\end{equation}

\begin{equation}
n_f = 6: \; \; \; \gamma_4 = 12.1820
\end{equation}
One can obtain a solution for $m[x(\mu)]$ which includes the ${\cal
O}(x^4)$ Pad\'e improvement of $\beta(x)$ and $\gamma(x)$ by
expressing $\gamma(x)/\beta(x)$ as a Maclaurin series in $x$, using
APAP estimates(3.3-6) for $\beta_4$ and (4.3-6) for $\gamma_4$. By
truncating this series after $x^4$, the differential equation (4.1)
can be approximated by

\begin{equation}
\frac{x}{m} \frac{dm}{dx} = \left[ \beta_0^{-1} + d_1 x + d_2 x^2 +
d_3 x^3 + d_4 x^4 \right],
\end{equation}
with $\beta_0$ given by (3.2a), and with $d_i$ given as follows:
\begin{equation}
n_f = 3:\; \; \; d_1 = 0.895063, \; d_2 = 1.94172, \; d_3 = 2.88956,
\; d_4 = \underline{417.493};
\end{equation}
\begin{equation}
n_f = 4:\; \; \; d_1 = 1.014131, \; d_2 = 1.74994, \; d_3 = 0.0880435,
\; d_4 = \underline{-3.93256};
\end{equation}
\begin{equation}
n_f = 5:\; \; \; d_1 = 1.17549, \; d_2 = 0.809817, \; d_3 = -1.05016,
\; d_4 = \underline{-10.0138};
\end{equation}
\begin{equation}
n_f = 6:\; \; \; d_1 = 1.39796, \; d_2 = 1.63266, \; d_3 = -6.84005,
\; d_4 = \underline{142.769};
\end{equation}
The values for $d_1, d_2$ and $d_3$ are exactly determined by
four-loop calculations of $\beta(x)$ and $\gamma(x)$.  The value of
$d_4$ is
underlined to emphasize that it is determined from APAP estimates.  
The large values for
$d_4$ when $n_f = 3$ and $n_f = 6$ reflect correspondingly large
values for $\beta_4$ that are discussed in the previous section.

The solution to (4.7) can be expressed in terms of $x(\mu)$ evaluated
at two different values of $\mu$:  $x(\mu_1) \equiv x_1$, $x(\mu_2)
\equiv x_2$, where $x(\mu)$ is the running coupling whose evaluation
is discussed in the previous section.  This solution to (4.7) is [4]

\renewcommand{\theequation}{4.12\alph{equation}}
\setcounter{equation}{0}

\begin{equation}
m(x_2) = m(x_1) c(x_2) / c(x_1),
\end{equation}
where
\begin{eqnarray}
c(x) & = & x^{1/\beta_0} \left\{ 1 + d_1 x + \left[ (d_1^2 + d_2) / 2
\right] x^2 \right. \nonumber \\
& + & \left[ (d_1^3 + 2d_3 + 3d_1 d_2) / 6 \right] x^3 \nonumber \\
& + & \left. \left[ (d_1^4 + 3d_2^2 + 6d_4 + 6d_1^2 d_2 + 8d_1 d_3) / 24
\right] x^4 \right\}.
\end{eqnarray}

Coefficients of $x$, $x^2$, and $x^3$ are determined in full by known
coefficients in the four-loop $\beta$ and $\gamma$ functions.  The
$x^4$ term is the lowest-degree term sensitive to Pad\'e driven
estimates of $\beta_4$ and $\gamma_4$.  We find the following set of
expressions for $c(x)$ from the $d_i$ in (4.8-11):

\renewcommand{\theequation}{4.1\arabic{equation}}
\setcounter{equation}{2}

\begin{eqnarray}
n_f = 3: \; \; \; c(x) & = & x^{4/9} \left[ 1 + 0.895063x + 1.37143 x^2
\right. \nonumber \\
& + & \left. 1.95168 x^3 + \underline{106.122 x^4} \right]
\end{eqnarray}
\begin{eqnarray}
n_f = 4: \; \; \; c(x) & = & x^{12/25} \left[ 1 + 1.01413x + 1.38920 x^2
\right. \nonumber \\
& + & \left. 1.09052 x^3 - \underline{0.0765827 x^4} \right] ,
\end{eqnarray}
\begin{eqnarray}
n_f = 5: \; \; \; c(x) & = & x^{12/23} \left[ 1 + 1.17549x + 1.50071 x^2
\right. \nonumber \\
& + & \left. 0.172486 x^3 - \underline{10.2813 x^4} \right] ,
\end{eqnarray}
\begin{eqnarray}
n_f = 6: \; \; \; c(x) & = & x^{4/7} \left[ 1 + 1.39796x + 1.79347 x^2
\right. \nonumber \\
& - & \left. 0.683486 x^3 + \underline{33.7949 x^4} \right] .
\end{eqnarray}
These same expressions are obtained to ${\cal O}(x^3)$ in ref. [4]; 
the effects of Pad\'e-improvement reside entirely in the $x^4$ terms.

It is important to recognise that these results ultimately derive
from applying the asymptotic error formula (2.3) to the perturbative
field-theoretical calculation of $\beta(x)$ and $\gamma(x)$, as
argued in [2] and [3].  As in the previous section, the results
(4.3-6) differ from those one would obtain using the fits of ref. [3]
to the coefficients of $n_f^k$ within $\gamma_4$, particularly as
$\gamma_4$ so extracted involves the near cancellation of large terms
from successive values of $k$:  from Table X of ref. [3] one finds for
$n_f = 5$ that $\gamma_4 = 530 - (143) \cdot 5 + (6.67) \cdot 5^2 +
(0.037) \cdot 5^3 - (8.54 \cdot 10^{-5}) \cdot 5^{4} = -13.7$. 
Small variations in these Pad\'e-estimated coefficients can easily
lead to positive values comparable to (4.5).

It is also important to note that the application of the APAP
algorithm at the field-theoretical level --- i.e., to $\beta(x)$ and
$\gamma(x)$ --- is {\it not} equivalent to applying it to
``perturbative'' expressions which are obtained by integrating over
these functions.  One could question, for example, whether the $x^4$
terms appearing in (4.13-16) might be obtainable by direct
application of the APAP algorithm (2.12) to lower-degree terms in
$x$. If we apply (2.12) directly to (4.13-16) using the explicit
coefficients of $x$, $x^2$ and $x^3$ in order to estimate the
coefficients of $x^4$, the $x^4$-coefficients we obtain are very
different from those listed.  Instead we obtain respectively for $n_f
= \{3,4,5,6\}: \; 2.683 x^4, 0.7426 x^4,
0.01839 x^4$, and $0.2850 x^4$.  This
discrepancy is indicative of the inapplicability of the error formula
(2.3) to the series in (4.12b), assuming (2.3) {\it is}
applicable to the perturbative field-theoretical series (3.1a) and
(4.2a). Such applicability is suggested by the predictions of
five-loop terms for $O(N)$ $\beta$- and $\gamma$-functions already
noted in Section 2.

Figure 3a-d display the relative impact on running quark masses 
of higher order corrections both
augmented and unaugmented by Pad\'e-improvement.  Given an initial
value $m_b(4.3
\; GeV) = 4.3 \; GeV$ [8], we evolve $m_b(\mu)$ up to $\mu=175$ GeV.  Fig
3a indicates the evolution obtained via (4.1) from three-loop
$\beta$- and $\gamma$-functions --- i.e. from truncation of the
series (3.1a) and (4.2a) after $i = 2$. Fig. 3b displays the relative effects of higher-order
corrections
augmented and unaugmented by Pad\'e-improvement.  We first consider
the unaugmented 4-loop case.  The upper curve in Fig 3b is the ratio
of $m_b$ obtained from 4-loop $\beta$- and $\gamma$-functions to
$m_b$ obtained from 3-loop $\beta$- and $\gamma$-functions (Fig 3a). 
As $\mu$ increases from 4 GeV to 175 GeV, the relative decrease in
$m_b(\mu)$ from use of four-loop information is seen to be less than
0.1\%.

Pad\'e improvement of four-loop results is displayed in the lower
curve of Fig 3b.  This curve is the ratio of a fully Pad\'e-improved
estimate of $m_b(\mu)$ to the three-loop calculation of $m_b(\mu)$
displayed in Fig 3a.  The full Pad\'e-improvement is obtained via an
APAP-algorithm determination of $\beta_4$ and $\gamma_4$, which is
then utilized to construct Pad\'e-summation [2$\vert$2] approximants as 
estimates of the aggregate effect of {\it all} higher-order terms 
in $\beta$ and $\gamma$.  The
[2$\vert$2] approximant for $\beta(x)$ is given by (3.7c);  the [2$\vert$2]
approximant within 
\begin{equation}
\gamma(x) = - x \left[ \frac{1 + 1.19485 x + 1.02765 x^2}{1 - 2.31904
x + 1.75664 x^2} \right]
\end{equation}
is obtained via (2.15-19) using the $n_f = 5$ values of $\gamma_1$,
$\gamma_2$, and $\gamma_3$ as well as the APAP-algorithm value for
$\gamma_4$ given in (4.5). The full [2$\vert$2] approximants are then
used to evaluate $c(x)$ in (4.12a):  $c(x) = \exp \left[ \int
\left( \gamma(x) / \beta(x) \right) dx \right]$.
It is evident from Fig 3b that
Pad\'e-improvement does not significantly alter $m_b(\mu)$ beyond a
correction comparable to five-loop expectations;  the relative change
from such Pad\'e improvement is respectively within 0.1\% (Fig 3b)
and 0.01\% of the unimproved three- and four-loop results.

Fig. 3c displays the corresponding evolution of the charmed quark
mass $m_c(\mu)$ for 1.3 GeV $\leq \mu \leq $  20 GeV, as obtained
from $\beta$- and $\gamma$-functions evaluated to two-, three-, and
four-loop order as well as from Pad\'e-improvement of the four-loop
$\beta$- and $\gamma$-functions, both below and above the
five-flavour threshold.  The initial value is taken to be $m_c$(1.3
GeV) = 1.3 GeV for all four curves, and the five-flavour threshold is
assumed to occur at 4.3 GeV [8].  The ``Pad\'e'' curve is obtained by
direct substitution of appropriate [2$\vert$2] Pad\'e-summation
$\beta$- and $\gamma$-functions into (4.1).  At the five-flavour
threshold, we utilize the threshold-matching constraint [9]
\begin{eqnarray}
m_q^{(n_f)} (\mu_{th}) & = & m_q^{(n_f-1)} (\mu_{th}) \left[ 1 +
0.2060 \left( \alpha_s^{(n_f)} (\mu_{th}) / \pi \right)^2 \right.
\nonumber \\
& + & \left. (1.8229 + 0.0247 n_f) (\alpha_s^{(n_f)} (\mu_{th}) /
\pi)^3 \right]^{-1}
\end{eqnarray}
with $n_f = 5$ and $\mu_{th} = 4.3$ GeV to generate the above-threshold
initial condition for $m_c$.  Thus, the above-threshold portion of
the Pad\'e curve in Fig. 3c is obtained from this initial condition
via substitution of (3.7c) and (4.17) into the differential
equation (4.1).  The below threshold portion of the Pad\'e curve is
obtained from the initial condition $m_c(1.3 \; GeV) = 1.3 \; GeV$
via substitution of the Pad\'e-summation four-flavour $\beta$-function
(3.7b) and four-flavour $\gamma$-function
\begin{equation}
\gamma(x) = - x \frac{[1 - 0.4541x - 1.3454 x^2]}{[1 - 4.1069x
+ 3.7090 x^2]}
\end{equation}
into the differential equation (4.1). The constraint (4.18) is also utilized to
generate the four-loop curve in Fig. 3c, and [when taken to
order-$\left( \alpha_s^{(n_f)}(\mu_{th}) \right)^2$] to generate the
three-loop curve.  As in Fig. 1, curves of successive order appear to
converge (at least qualitatively) to the Pad\'e estimate, which is
almost indistinguishable from the four-loop curve.

\renewcommand{\theequation}{4.\arabic{equation}}
\setcounter{equation}{19}

Pad\'e-improvement effects are somewhat larger for light quarks (u,
d, s).  The evolution of light quarks from an initial value
(normalized to unity) at $\mu$ = 1 GeV is displayed in Fig. 3d. This
latter set of curves is obtained via utilization of (4.18) at both
$n_f = 4$ and $n_f = 5$ flavour thresholds, which are respectively
taken to be at 1.3 GeV and 4.3 GeV [8].
Below the 1.3 GeV four-flavour threshold, the Pad\'e curve is
generated via (4.1) using the $n_f = 3$ Pad\'e-summation
$\gamma$-function
\begin{equation}
\gamma(x) = -x \frac{[1 - 1.2373x - 1.8485x^2]}{[1 - 5.0289x + 4.7993
x^2]}
\end{equation}
and the $n_f = 3$ $\beta$-function (3.7a).  Between the four- and
five-flavour thresholds we utilize (4.19) and (3.7b), and above the
five-flavour threshold, we utilize (4.17) and (3.7c) as before.
At $\mu = 5$ GeV, Fig. 3d 
shows that there is a 1\% difference between running masses obtained via
unimproved and Pad\'e-improved four-loop $\beta$- and
$\gamma$-functions.  Once again, however, the distance between curves
of successive order decreases as the order increases, giving 
the appearance of convergence towards the
Pad\'e-improved curve.

\section*{5.  Application to Higgs Decays}

Although the Higgs particle has yet to be directly observed,
expressions for its decay into either two-gluons [10] or a $b \bar{b}$
pair [11] have been worked out with precision in perturbation theory. 
In much the same way, knowledge of the $Z^\circ$ decay widths, whose
precise values are important for bounds on standard-model parameters,
preceded the discovery of the $Z^\circ$ itself.  In this section, we
apply Pad\'e-improvement to the decay processes $H \rightarrow$ two
gluons and $H \rightarrow$ $b \bar{b}$, and examine whether such
improvement leads to detectable changes from the calculated rates
obtained without such Pad\'e improvement.

\subsection*{5.1  Higgs $\rightarrow$ Two Gluons}

The decay rate $H \rightarrow gg$ has been calculated to three-loop
order in perturbation theory [10]:

\renewcommand{\theequation}{5.\arabic{equation}}
\setcounter{equation}{0}

\begin{eqnarray}
\Gamma (H \rightarrow gg) & = & \frac{G_F M_H^3 x_H^2}{36 \pi \sqrt{2}} \times
\left[ 1 + 17.9167 x_H \right. \nonumber \\
& + & \left. \left( 156.808 - 5.70833 \; ln (m_t^2 / M_H^2) \right)
x_H^2 + {\cal O}(x_H^3) \right],
\end{eqnarray}
where $x_H = x(M_H) = \alpha_s (M_H) / \pi$, and where $M_H$ is
assumed to be less than $m_t$.  Pad\'e-improvement can enter this
expression both in the actual value of $\alpha_s(M_H)$ evolving from
a Pad\'e-improved $\beta$-function, as well as in a Pad\'e-driven
estimate of the ${\cal O} (x_H^3)$ contribution to the square
bracketed expression in (5.1).  One cannot apply the APAP-algorithm
of Section 2 to estimate this term because for a given value of
$M_H$, only the coefficient of $x$ $(R_1 = 17.9167)$ and $x^2 \left[ R_2
= 156.808 - 5.70833 \; ln (m_t^2 / M_H^2) \right]$ are known;  the
coefficient $R_3$ of $x^3$ is {\it not} known.  One way to
estimate $R_3$ is to express $1 + R_1 x + R_2 x^2$ as a [1$\vert$1] Pad\'e
approximant, which upon expansion yields $R_3 = R_2^2 / R_1$.  A
refinement on this estimate that is actually utilized to approximate
$\beta_3$ in ref. [2] is to assume in the asymptotic error formula
(2.3) that $a + b$ is small compared to $N + 1$.  One can then argue
from (2.5) and (2.7) that $\delta_3 \simeq \delta_2 / 2$, where
$\delta_2 = (R_1^2 - R_2) / R_2$ is determined in full by $R_1$ and
$R_2$.  Eq. (2.7) can then be rearranged to yield the following
estimate of $R_3$:
\begin{equation}
R_3 = \frac{R_2^2/R_1}{1 + \delta_3} = \frac{2 R_2^3}{R_1^3 + R_1
R_2}.
\end{equation}

For different $M_H$ values we plot in Fig. 4 the ratio of the
Pad\'e-improved $H \rightarrow gg$ rate to the rate obtained directly
from (5.1), using the three-loop $\beta$ function to obtain $x_H$
from the initial condition $x(M_z) = 0.118/\pi$ [8].  The
Pad\'e-improvement of the $H \rightarrow gg$ rate is obtained for a
given choice of $M_H$ first by evolving $x_H$ from the same initial
condition via the [2$\vert$2] approximant (3.7c) for the $\beta$-function,
then by using (5.2) to estimate the ${\cal O}(x_H^3)$ term in (5.1),
and finally by replacing the now-known cubic $1 + R_1 x_H + R_2 x_H^2
+ R_3 x_H^3$ in (5.1) with its appropriate [2$\vert$1] Pad\'e summation:
\begin{equation}
1 + R_1 x + R_2 x^2 + R_3 x^3 \rightarrow \frac{1 + (R_1 - R_3 / R_2) x 
+ (R_2 - R_3 R_1 / R_2)x^2}{1 - (R_3 / R_2) x}.
\end{equation}

Figure 4 shows that such Pad\'e improvement yields a 2.5\% - 3\%
increase in the $H \rightarrow gg$ rate, with very little sensitivity
to the Higgs mass.  Such improvement is best understood to be an
estimate of (unknown) higher order corrections to (5.1) that should
be eventually testable against both experimental and future
higher-order calculations of the $H \rightarrow gg$ rate.

\subsection*{5.2 Higgs $\rightarrow b \bar{b}$}

The decay rate Higgs $\rightarrow b \bar{b}$ has been calculated to
four-loop order in perturbation theory [11]:
\begin{eqnarray}
\Gamma (H \rightarrow b \bar{b}) = \left[ 3 G_F M_H m_b^2 (M_H) / (4
\pi \sqrt{2}) \right] \nonumber \\
\times \left\{ \left[ 1 + (17 / 3) x_H + 29.1467 x_H^2 + 41.7581 x_H^3
\right] \right. \nonumber \\
- \left. (6 m_b^2 (M_H) / M_H^2) [1 + (20/3) x_H + 14.62 x_H^2]
\right\}
\end{eqnarray}
Full Pad\'e-improvement of this expression for a given value of $M_H$
(with $M_H < m_t$) entails...

\begin{description}
\item 1)  ...determination of $x(M_H)$ through use of (3.7), the
[2$\vert$2] Pad\'e-summation of the $\beta$-function, to evolve $x(\mu)$
from an appropriate initial condition; {\it e.g.} $\alpha_s (M_z) =
0.118$ [8];

\item  2)  ...determination of $m_b(M_H)$ through substitution into
(4.1) of (3.7c) and (4.17), the [2$\vert$2] Pad\'e summations for
$\beta(x)$ and $\gamma(x)$, so as to evolve $m_b(\mu)$ from an
appropriate initial condition; {\it e.g.} $m_b(4.3 GeV) = 4.3 \; GeV$ [8];

\item 3)  ...Pad\'e-improvement and [2$\vert$2] Pad\'e summation of the
cubic expression in (5.4), 
\begin{equation}
1 + (17/3) x + 29.1467 x^2 + 41.7581 x^3 \rightarrow \frac{[1 + 4.30262 x +
21.0641 x^2]}{[1 - 1.36405 x - 0.352971 x^2]},
\end{equation}
where (2.12) is used to generate an estimate of the $x^4$ coefficient
[67.2472], and where (2.15 - 19) are used to generate the [2$\vert$2]
approximant in (5.5), and 

\item 4)  ...Pad\'e-improvement and [2$\vert$1] Pad\'e summation of the
quadratic expression in (5.4),
\begin{equation}
1 + (20/3) x + 14.62 x^2 \rightarrow \frac{[1 + 5.581 x + 7.382 x^2]}{[1 -
1.086 x]}
\end{equation}
where (5.2) is used to generate an estimate of the $x^3$ coefficient
[15.87], and where (5.3) is used to generate the [2$\vert$1] approximant in
(5.6).
\end{description}

The relative size of all these corrections, referenced to the rate
calculated to the next-to-highest-known [three-loop] order in
perturbation theory, is displayed in Fig. 5.  The top curve is the
ratio $[\Gamma(H \rightarrow b \bar{b})]_{4-loop} / [\Gamma(H
\rightarrow b \bar{b})]_{3-loop}$, as a function of $M_H$.  The
4-loop rate is obtained directly from (5.4), with $m_b(M_H)$ and
$x_H(= x(M_H))$ obtained from $\beta$- and $\gamma$-functions that
are truncated to zero after their $\beta_3$ and $\gamma_3$
contributions.  The 3-loop rate is obtained by truncating off 
the highest order terms in (5.4) --- specifically the
${\cal O}(x^3)$ term on the left-hand side of (5.5) and the ${\cal
O}(x^2)$ term on the left-hand side of (5.6) --  and by obtaining
$x_H$ and $m_b(M_H)$ from $\beta$- and $\gamma$-functions that are
truncated to zero after $\beta_2$ and $\gamma_2$ contributions.  The
top curve shows a change of 0.02\% to 0.09\% in going fom three to
four-loop order.

The bottom curve compares the ratio of $\Gamma(H \rightarrow b
\bar{b})$, obtained by full Pad\'e improvement of $[\Gamma(H
\rightarrow b \bar{b})]$, as described above, to $[\Gamma(H
\rightarrow b \bar{b})]_{3-loop}$.  It is evident from the figure
that the two ratios are within 0.0001 of each other for all values of
$M_H$ below $m_t$.  In other words, Pad\'e-improvement reduces the
4-loop result for $\Gamma(H \rightarrow b \bar{b})$ by {\it at
most} 0.01\%.  The effect that {\it is} seen seems to derive
wholly from the Pad\'e improvement of $m_b(\mu)$, as is evident from
comparison of Fig 5 to Fig 3b.  Of course, such close agreement
between (5.4) and its fully Pad\'e-improved version suggests that the
expression (5.4) is more than adequate for future comparison to
experiment.  Thus the purpose of the analysis presented here 
is really
to demonstrate the robustness of (5.4) against Pad\'e estimates of
higher order corrections.

The small size of corrections past even the 3-loop order is partly
consequence of the small size of $x(M_H)$ characterizing Higgs decay
rates.  Pad\'e corrections are of much more interest when the
magnitude of $x$ is larger, suggesting their usefulness in assessing the
perturbative content of low-energy QCD -- {\it i.e.} QCD sum rules.  In the
section that follows we will address how Pad\'e-improvement can be
utilized to estimate substantial higher-order corrections to sum
rules relevant to scalar- and pseudoscalar-meson static properties.

\section*{6.  Perturbative Content of Scalar/Pseudoscalar QCD Sum Rules}

\renewcommand{\theequation}{6.\arabic{equation}}
\setcounter{equation}{0}

The resonance content of finite-energy ${\cal F}_k$ and Laplace
${\cal R}_k$ QCD sum rules [12,13] is obtained from integrals over
the imaginary part of current correlation functions $\Pi(s, \mu^2)$
in the subcontinuum region $(s < s_0)$:

\begin{equation}
{\cal F}_k (s_0) = \frac{1}{\pi} \int_0^{s_0} Im \Pi (s, s_0) s^k ds,
\end{equation}
\begin{equation}
{\cal R}_k (\tau, s_0) = \frac{1}{\pi} \int_0^{s_0} Im \Pi(s,
1/\tau)) s^k e^{-s \tau} ds.
\end{equation}
We consider here the purely perturbative content of the correlation
function for scalar currents, which is presently known to 4-loop
order [11]:
\begin{eqnarray}
& & \frac{1}{\pi} Im \Pi (s, \mu^2) \nonumber \\
& = & \frac{3s}{8 \pi^2} \left\{ 1 + \left( \frac{\alpha_s(\mu)}{\pi}
\right) \left[ a_0 + a_1 ln \left( \frac{\mu^2}{s} \right) \right]
\right. \nonumber \\
& + & \left( \frac{\alpha_s(\mu)}{\pi}\right)^2 \left[ b_0 + b_1 ln
\left( \frac{\mu^2}{s} \right) + b_2 \left( ln \left( \frac{\mu^2}{s}
\right) \right)^2 \right] \nonumber \\
& + & \left( \frac{\alpha_s (\mu)}{\pi} \right)^3 \left[ c_0 + c_1 ln
\left( \frac{\mu^2}{s} \right) + c_2 \left( ln \left( \frac{\mu^2}{s}
\right) \right)^2 + c_3 \left( ln \left( \frac{\mu^2}{s} \right)
\right)^3 \right] \nonumber \\
& + & \left. \left( \frac{\alpha_s(\mu)}{\pi} \right)^4 R_4 + ...
\right\}.
\end{eqnarray}
The problem we will address in this section is the computation of
$R_4$, which is necessary for the determination of ${\cal
O}(\alpha_s^4)$ contributions to ${\cal F}_k$ and ${\cal R}_k$.  For
three flavours Chetyrkin [11] has found that

\renewcommand{\theequation}{6.4\alph{equation}}
\setcounter{equation}{0}
 
\begin{equation}
a_o = 17/3, \; \; a_1 = 2,
\end{equation}
\begin{equation}
b_0 = 31.8640, \; \; b_1 = 31.6667, \; \; b_2 = 17/4,
\end{equation}
\begin{equation}
c_0 = 89.1564, \; \; c_1 = 297.596, \; \; c_2 = 229/2, \; \; c_3 =
9.20833.
\end{equation}
If we define $w \equiv s/\mu^2$, we can estimate $R_4[w]$ directly by
substituting

\renewcommand{\theequation}{6.5\alph{equation}}
\setcounter{equation}{0}

\begin{equation}
R_1[w] = a_0 - a_1 \; ln \; w
\end{equation}

\begin{equation}
R_2[w] = b_0 - b_1 \; ln \; w + b_2 (ln \; w)^2
\end{equation}

\begin{equation}
R_3[w] = c_0 - c_1 \; ln \; w + c_2 (ln \; w)^2 - c_3 (ln \; w)^3
\end{equation}
directly into (2.12).  This enables one to determine explicitly
${\cal O}(\alpha_s^4)$ corrections to the sum-rules (6.1) and (6.2), even
though $R_4[w]$ determined in this way is manifestly not a
fourth-order polynomial in $ln(w)$.  In particular, we easily find
the ${\cal O}(\alpha_s^4)$ contribution to ${\cal F}_k(s_0)$ to be

\renewcommand{\theequation}{6.6\alph{equation}}
\setcounter{equation}{0}

\begin{eqnarray}
\Delta{\cal F}_0 (s_0) & = & \frac{3 s_0^2}{16 \pi^2} \left(
\frac{\alpha_s (s_0^{1/2})}{\pi} \right)^4 \int_0^1 2 R_4 [w] \; dw
\nonumber \\
& = & \frac{3 s_0^2}{16 \pi^2} \left( \frac{\alpha_s(s_0^{1/2})}{\pi}
\right)^4 (2059.4),
\end{eqnarray}

\begin{eqnarray}
\Delta{\cal F}_1 (s_0) & = & \frac{s_0^3}{8 \pi^2} \left(
\frac{\alpha_s (s_0^{1/2})}{\pi} \right)^4 \int_0^1 3 R_4 [w] \; w \;dw
\nonumber \\
& = & \frac{s_0^3}{8 \pi^2} \left( \frac{\alpha_s(s_0^{1/2})}{\pi}
\right)^4 (1158.4),
\end{eqnarray}

\begin{eqnarray}
\Delta{\cal F}_2 (s_0) & = & \frac{3 s_0^4}{32 \pi^2} \left(
\frac{\alpha_s (s_0^{1/2})}{\pi} \right)^4 \int_0^1 4 R_4 [w] \; w^2 \;dw
\nonumber \\
& = & \frac{3 s_0^4}{32 \pi^2} \left( \frac{\alpha_s(s_0^{1/2})}{\pi}
\right)^4 (833.47).
\end{eqnarray}
The integrals in (6.6) have been evaluated numerically.  
This approach, however, ignores the known
structural dependence of $R_4$ on the variable $w$,

\renewcommand{\theequation}{6.\arabic{equation}}
\setcounter{equation}{6}

\begin{equation}
R_4 [w] = d_0 - d_1 ln \; w + d_2 (ln \; w)^2 - d_3 (ln \; w)^3 + d_4
(ln \; w)^4,
\end{equation}
which may be important when one integrates over the $w$-variable, as
in (6.6).  The ${\cal O}(\alpha_s^4)$ corrections to the first three
finite energy sum rules are easily determined in terms of the
constants $d_i$ by substitution of (6.7) into the integrand of (6.1):

\begin{equation}
\Delta {\cal F}_0 (s_0) = \frac{3 s_0^2}{16 \pi^2} \left(
\frac{\alpha_s (s_0^{1/2})}{\pi} \right)^4 \left( d_0 + \frac{d_1}{2} +
\frac{d_2}{2} + \frac{3d_3}{4} + \frac{3d_4}{2} \right),
\end{equation}

\begin{equation}
\Delta {\cal F}_1 (s_0) = \frac{s_0^3}{8 \pi^2} \left(
\frac{\alpha_s (s_0^{1/2})}{\pi} \right)^4 \left( d_0 + \frac{d_1}{3} +
\frac{2d_2}{9} + \frac{2d_3}{9} + \frac{8d_4}{27} \right),
\end{equation}

\begin{equation}
\Delta {\cal F}_2 (s_0) = \frac{3 s_0^4}{32 \pi^2} \left(
\frac{\alpha_s (s_0^{1/2})}{\pi} \right)^4 \left( d_0 + \frac{d_1}{4} +
\frac{d_2}{8} + \frac{3d_3}{32} + \frac{3d_4}{32} \right),
\end{equation}

We can use the Pad\'e algorithm (2.12) to estimate the coefficients
$d_i$.  To do so, we let $R_1$, $R_2$, and $R_3$ be given by (6.5)
for five representative values of $w$ between zero and one:
$w = \left\{ 1, e^{-1/4}, e^{-1/2}, e^{-1}, e^{-2} \right\}$.  When
$w = 1$, corresponding to $s = s_0$ in the finite-energy sum rule
integrand (6.1), we see from (6.5) that $R_1 = a_0$, $R_2 = b_0$,
$R_3 = c_0$.  Using the APAP-algorithm (2.12), we find that $R_4[1] =
251.442 = d_0$.  When $w = e^{-1/4} \; (s = 0.779 s_0)$, we find from
(6.5) that $R_1 = 37/6$, $R_2 = 40.0463$, $R_3 = 170.8554$.  Using
(2.12) and (6.7), respectively, we then find that
\begin{equation}
R_4[e^{-1/4}] = 699.398 = d_0 + d_1 / 4 + d_2 / 16 + d_3 / 64 + d_4 /
256.
\end{equation}
Similarly we find the following results when $w = e^{-1/2} \; (s = 0.606
s_0)$, $w = e^{-1} \; (s = 0.368 s_0)$, and $w = e^{-2} \; (s = 0.135
s_0)$:
\begin{equation}
R_4[e^{-1/2}] = 1389.82 = d_0 + d_1 / 2 + d_2 / 4 + d_3 / 8 + d_4 /
16,
\end{equation}
\begin{equation}
R_4[e^{-1}] = 3652.36 = d_0 + d_1 + d_2 + d_3 + d_4,
\end{equation}
\begin{equation}
R_4[e^{-2}] = 12804.9 = d_0 + 2d_1 + 4d_2 + 8d_3 + 16d_4.
\end{equation}
We solve the four linear equations (6.11-14) for the four unknowns
$d_1, d_2, d_3, d_4$ using the value already obtained for $d_0 ( =
251.422)$, and we obtain $d_1 = 1357.84, \; d_2 = 1634.53, \; d_3 = 404.630, \;
d_4 = 3.9097$. Substitution of these numbers into (6.8-10) yields
results remarkably close to those obtained in (6.6).  These results
are listed in the underlined highest-order terms given below for the
Pad\'e-improved perturbative content of the first three finite energy
sum rules $(x \equiv  \alpha_s (s_0^{1/2}) / \pi)$:

\renewcommand{\theequation}{6.15\alph{equation}}
\setcounter{equation}{0}

\begin{eqnarray}
{\cal F}_0(s_0) & = &  \frac{3 s_0^2}{16 \pi^2} \left[ 1 + 6.66667x +
49.8223x^2 \right. \nonumber \\
& + & \left. 302.110x^3 + \underline{2057.0}x^4 \right],
\end{eqnarray}
\begin{eqnarray}
{\cal F}_1(s_0) & = &  \frac{s_0^3}{8 \pi^2} \left[ 1 + 6.33333x +
43.3640x^2 \right. \nonumber \\
& + & \left. 215.846x^3 + \underline{1158.4}x^4 \right],
\end{eqnarray}
\begin{eqnarray}
{\cal F}_2(s_0) & = &  \frac{3 s_0^4}{32 \pi^2} \left[ 1 + 6.16667x +
40.3119x^2 \right. \nonumber \\
& + & \left. 178.731x^3 + \underline{833.52}x^4 \right].
\end{eqnarray}
Pad\'e corrections to the Laplace sum rules (6.2) are not listed, as
they are complicated by the occurrence of two scale variables ($s_0$ and
$\tau$).  However, such corrections are straightforward to obtain
via integration of the ${\cal O}(\alpha_s^4)$ term of (6.3), which we
have already obtained via APAP estimates of $d_{0-4}$:

\renewcommand{\theequation}{6.1\arabic{equation}}
\setcounter{equation}{5}

\begin{eqnarray}
R_4 & = & 251.44 + 1357.8 \; ln \left( \frac{\mu^2}{s}
\right) + 1634.5 \left( ln \left( \frac{\mu^2}{s} \right) \right)^2
\nonumber \\
& + & 404.63 \left( ln \left( \frac{\mu^2}{s} \right) \right)^3 + 3.9097
\left( ln \left( \frac{\mu^2}{s} \right) \right)^4.
\end{eqnarray}

It is worth noting that the underlined $R_4$ terms in (6.15) are not very
different from those one would obtain using either the $R_3^2 / R_2$
estimate suggested by a [2$\vert$1] approximant, as discussed following
(2.10), or the APAP algorithm (2.12) applied directly to the known
(non-underlined) terms of (6.15).  The increase in the size of
coefficients with increasing powers of $x$ suggests the utility of a
[2$\vert$2] Pad\'e summation for these three expressions as an improvement
over truncating off what may be substantial ${\cal O}(x^5)$
corrections to (6.15).  By applying (2.15-19) to the three equations
(6.15), we obtain the following [2$\vert$2] approximants to the full
perturbative content of the first three finite-energy sum rules:

\renewcommand{\theequation}{6.17\alph{equation}}
\setcounter{equation}{0}

\begin{equation}
{\cal F}_0(s_0) = \frac{3 s_0^2}{16 \pi^2} \left[ \frac{1 + 3.8073x +
6.8124x^2}{1 - 2.8594x - 23.947x^2} \right],
\end{equation}
\begin{equation}
{\cal F}_1(s_0) = \frac{s_0^3}{8 \pi^2} \left[ \frac{1 + 2.3917x +
11.308x^2}{1 - 3.9416x - 7.0930x^2} \right],
\end{equation}
\begin{equation}
{\cal F}_2(s_0) = \frac{3 s_0^4}{32 \pi^2} \left[ \frac{1 + 2.2174x +
12.791x^2}{1 - 3.9492x - 3.1670x^2} \right].
\end{equation}

Both (6.15a) and (6.17a) are indicative of a need for $s_0^{1/2}$ to
be substantially larger than 1 GeV for finite-energy sum rules to be
useful in the scalar and pseudoscalar channels.  If $s_0^{1/2} = 1$ GeV, 
we see from Fig 1 that $x(1 GeV) = 0.153$.  For this value of
$x$, each successive term in the square brackets of (6.15a) is
approximately unity, indicative of nonconvergence.  This is reflected
by a near-vanishing of the denominator of (6.17a), implying a
divergent result for the summation of the full perturbative series. 
If $s_0 = 3.24 \; GeV^2$, we see from Fig. 1 that $x(s_0^{1/2}) \cong
0.10$.  The truncated series (6.15a) is then seen to yield a value
that is only 87\% of that obtained via the Pad\'e summation (6.17a),
indicative of the magnitude of the higher order terms missing from
(6.15a).  Note that a choice for $s_0$ near or somewhat above $3
\; GeV^2$ is suggested by Laplace sum-rule fits in both the pseudoscalar
[14] and scalar [15] resonance channels.

The finite energy sum rule ${\cal F}_0(s_0)$ provides an example of
how it is not enough just to have precise higher-order results.  Even
though (6.15a) includes 4-loop effects as well as an APAP-algorithm
estimate of 5-loop effects, the five terms listed demonstrate only
sluggish convergence for a realistic choice of $s_0$.  There is found
to be enough of a difference between the truncated series (6.15a) and
its Pad\'e-summation (6.17a) to suggest the advisability of using 
the latter.

\section*{7. Summary}

Using a Pad\'e-motivated algorithm (2.12), we have estimated in
Section 3 the five-loop contributions to the $\beta$-function 
for $n_f = \{3,4,5,6\}$, and
we have compared the evolution of $\alpha_s(\mu)$ from $\mu = M_z$ obtained from two-loop, 
three-loop, four-loop,
and Pad\'e-summation estimates of the full $\beta$-function.  Low 
energy values of $\alpha_s$  
obtained from the four-loop $\beta$-functions with quoted flavour thresholds and appropriate
threshold
matching conditions are within 1\% of those obtained from the Pad\'e-summation 
$\beta$-functions, a small effect compared to the much larger 
sensitivity of $\alpha_s$(1 GeV) 
and $\alpha_s(m_\tau)$ to present uncertainties in $\alpha_s(M_z)$ and c- and 
b-quark flavour thresholds.

We concluded Section 3 by extracting the most general set of [2$\vert$2] 
Pad\'e-summation estimates of 3, 4,
5, and 6 flavour QCD $\beta$-functions whose Maclaurin expansions yield known four-loop results for
their
first four terms.  For positive values of $\alpha_s$, these Pad\'e-summation 
estimates of the $\beta$-function were shown to alternate denominator and 
numerator zeros,
{\it regardless} of the size of the (presently unknown) five-loop term 
serving as a free parameter in the approximants.  Such alternation 
necessarily implies that all positive numerator zeros 
represent {\it ultraviolet} fixed points, behaviour which, if
applicable to the true $\beta$-function, would decouple the 
(suitably defined) infrared
region from perturbative QCD.

In Section 4, we applied Pad\'e-improvement methods to the running quark mass by
estimating five-loop contributions to the $\gamma$-functions for 3, 4, 5, and 6 quark flavours. 
We then
extracted an estimate for the $O(x^4)$ contribution to closed-form expressions 
for $m_q[\alpha_s(\mu)/\pi]$ that
had been obtained earlier [4,6] to $O(x^3)$. We compared the evolution of 
3-loop, 4-loop, and
Pad\'e-summation estimates of $m_b(\mu)$, once again finding very little 
relative difference (0.01\%)
between Pad\'e-summation and four-loop determinations of $m_b(\mu)$ over the 
range $\mu < m_t$, given
identical five-flavour-threshold initial conditions.  Corresponding
agreement was still seen to occur at the 1\% level for light quarks. 

In Section 5 we applied the results of the previous two sections to higher-loop calculations
of the Higgs decay ratio $H \rightarrow gg$ and $H \rightarrow
b\bar{b}$, rates which are sensitive to running couplings and
running masses, as well as higher-loop corrections that are polynomial 
in $\alpha_s(m_H)$.  The calculated
three-loop $H \rightarrow gg$ rate is shown to be within 3\% of our
Pad\'e-improvement estimate, given
identical choices for $M_H, \alpha(M_z)$, and $m_b$-threshold initial 
conditions.  Similarly, the calculated
four-loop $H \rightarrow b\bar{b}$ rate is seen to differ from full 
Pad\'e-improvement by at most 0.01\%.

All the results summarized up until this point are indicative of close agreement between
known perturbation theory and Pad\'e-approximant improvements
intended to
take into account higher-order effects.  Consequently, the theoretical uncertainties associated with
the truncation of any such calculations at the three-or-four-loop order are shown to be small.  In
Section 6, we considered quantities known to four-loop order for which this is
{\it not} the case, the
purely-perturbative content of QCD sum rules in
scalar/pseudoscalar-resonance channels.  We constructed a
Pad\'e-algorithm estimate of the
purely-perturbative $O(\alpha_s^4)$ contribution to the imaginary part of 
scalar/pseudoscalar correlation functions, and we obtained
[2$\vert$2] Pad\'e-summation estimates of the all-orders content of the first 
three finite energy sum rules.  We found the overall convergence 
of the primary sum rule to be doubtful
for values of the QCD continuum threshold near $s_0$ = 1 GeV$^2$.  Even for
$s_0$  above 3 GeV$^2$, we
found a greater-than-10\% discrepancy between Pad\'e-summation and 
four-loop-order contributions to this sum rule, suggesting the existence 
of substantial theoretical uncertainties from higher-than-four-loop 
contributions. 


\section*{Acknowledgements}

      We are grateful for Mark Samuel's participation in the early stages of this work, prior to
his sudden passing last November.  His enthusiasm, energy, and sharp intelligence will be missed
by all who had the opportunity to work with him.
We are also grateful to A.L. Kataev for a seminal discussion on higher-order terms in
sum rules, K.G. Chetyrkin for useful correspondence on both the running quark
mass and scalar correlation functions, and to the 
Natural Sciences and Engineering
Research Council of Canada for research support.
VE is grateful for the hospitality of the Department of Physics and Engineering
Physics at the University of Saskatchewan.

\newpage


\section*{References:}

\begin{enumerate}

\item{}M. A. Samuel, G. Li, and E. Steinfelds, Phys. Rev. D 48, 869 (1993) 
and Phys. Rev. E
      51, 3911 (1995); M. A. Samuel, J. Ellis, and M. Karliner, 
Phys. Rev. Lett. 74, 4380
      (1995).

\item{}J. Ellis, M. Karliner, and M. A. Samuel, Phys. Lett. B 400, 176 (1997).

\item{}J. Ellis, I. Jack, D. R. T. Jones, M. Karliner, and M. A. Samuel, 
      Phys. Rev. D 57, 2665 (1998).

\item{}K. G. Chetyrkin, Phys. Lett. B 404, 161 (1997).

\item{}T. van Ritbergen, J. A. M. Vermaseren, and S. A. Larin, Phys. Lett. B 400, 379 (1997).

\item{}T. van Ritbergen, J. A. M. Vermaseren, and S. A. Larin, 
      Phys. Lett. B 405, 323 (1997).

\item{}H. Kleinert, J. Neu, V. Schulte-Frohlinde, K. G. Chetyrkin, and S. A. Larin, Phys. Lett.
      B 272, 39 (1991) and B 319, 545 (1993) Erratum.

\item{}R. M. Barnett et al. [Particle Data Group], Phys. Rev. D 54, 1 (1996).

\item{}K. G. Chetyrkin, B. A. Kniehl, and M. Steinhauser, Max-Planck-Institut Preprint
      MPI/PhT/97-041 [hep-ph/9708255 v2].

\item{}K. G. Chetyrkin, B. A. Kniehl, and M. Steinhauser, Phys. Rev. Lett. 79, 353 (1997).

\item{}K. G. Chetyrkin, Phys. Lett. B 390, 309 (1997).

\item{}W. Hubschmid and S. Mallik, Nucl. Phys. B 193, 368 (1981).

\item{}M. A. Shifman, A. I. Vainshtein, and V. I. Zakharov, Nucl. Phys. B 147, 385 and 448
      (1979).

\item{}V. Elias, A. H. Fariborz, M. A. Samuel, Fang Shi, and T. G. Steele, Phys. Lett. B 412,
      131 (1997); T. G. Steele, J. C. Breckenridge, M. Benmerrouche, V. Elias, and A. H.
      Fariborz, Nucl. Phys. A 624, 517 (1997).

\item{}V. Elias, A. H. Fariborz, Fang Shi, and T. G. Steele, Nuc. Phys. A 633, 279 (1998).

\end{enumerate}

\newpage

\begin{table}
  \begin{center}
    \begin{tabular}{cccccc}
$$ & $\beta_4$ via (2.12) & $\beta_4^{true}$ & $\beta_4$ via [2] &
$(\gamma_4 / \gamma_0)$ via (2.12) & $(\gamma_4 / \gamma_0)^{true}$ \\
\hline
N = 1 & 1421.7 & 1424.3 & 1432.3 & 135.1 & 150.76 \\
N = 2 & 1941.7 & 1922.3 & 1943.8 & 168.4 & 191.89 \\
N = 3 & 2555.9 & 2499.3 & 2540.3 & 203.1 & 236.94 \\
N = 4 & 3267.9 & 3158.8 & 3225.6 & 239.2 & 285.94
\end{tabular}
\caption{Comparison of $O(N)$ $\beta$-function and $\gamma$-function
coefficients obtained via Pad\'e estimates to those obtained via exact calculation. 
The ``$\beta_4$ via (2.12)'' column displays estimates obtained via the APAP 
algorithm (2.12),
based on $\beta_{0-3}$ listed in [7].  The ``$\beta_4^{true}$'' column 
displays the results for $\beta_4$ obtained in [7] by explicit calculation. 
The ``$\beta_4$ via [2]'' column
lists the Pad\'e estimates given in Table 3 of ref. [2], involving
knowledge of the $N^4$ dependence and a fit of the overall
$N$-dependence of $\beta_4$, as well as a simplified asymptotic error
formula.  The ``$(\gamma_4 / \gamma_0)$ via (2.12)'' column displays
estimates of $\gamma_4 / \gamma_0$ in the anomalous mass dimension
$\gamma$-function obtained via the APAP algorithm (2.12), based on
prior coefficients listed in [7].  The final column,
``$(\gamma_4/\gamma_0)^{true}$'' are coefficients explicitly
calculated in [7].}
\end{center}
\end{table}

{\bf Figure Captions:}

\begin{description}

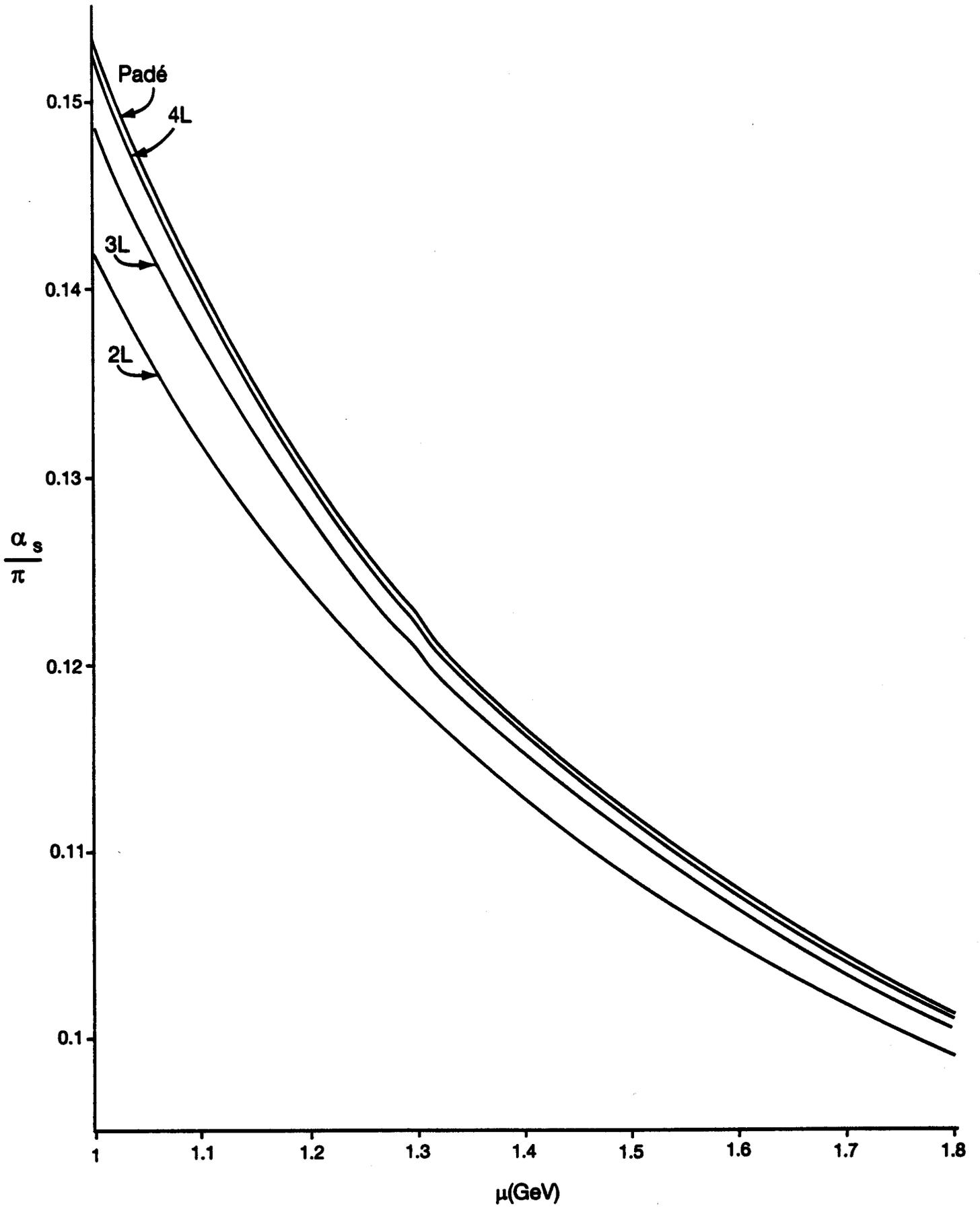
\item[Figure 1:] A comparison of the evolution of $x(\mu) \equiv
\alpha_s (\mu) / \pi$ for $m_t \geq \mu \geq 1$ GeV obtained from the
evolution of two-loop (2L), three-loop (3L), four-loop (4L), and
[2$\vert$2] Pad\'e-summation $\beta$-functions.  All curves are
generated from the initial condition $\alpha_s(M_z) = 0.118$, with
five- and four-flavour threshold matchings occurring at 4.3 GeV and
1.3 GeV, respectively.

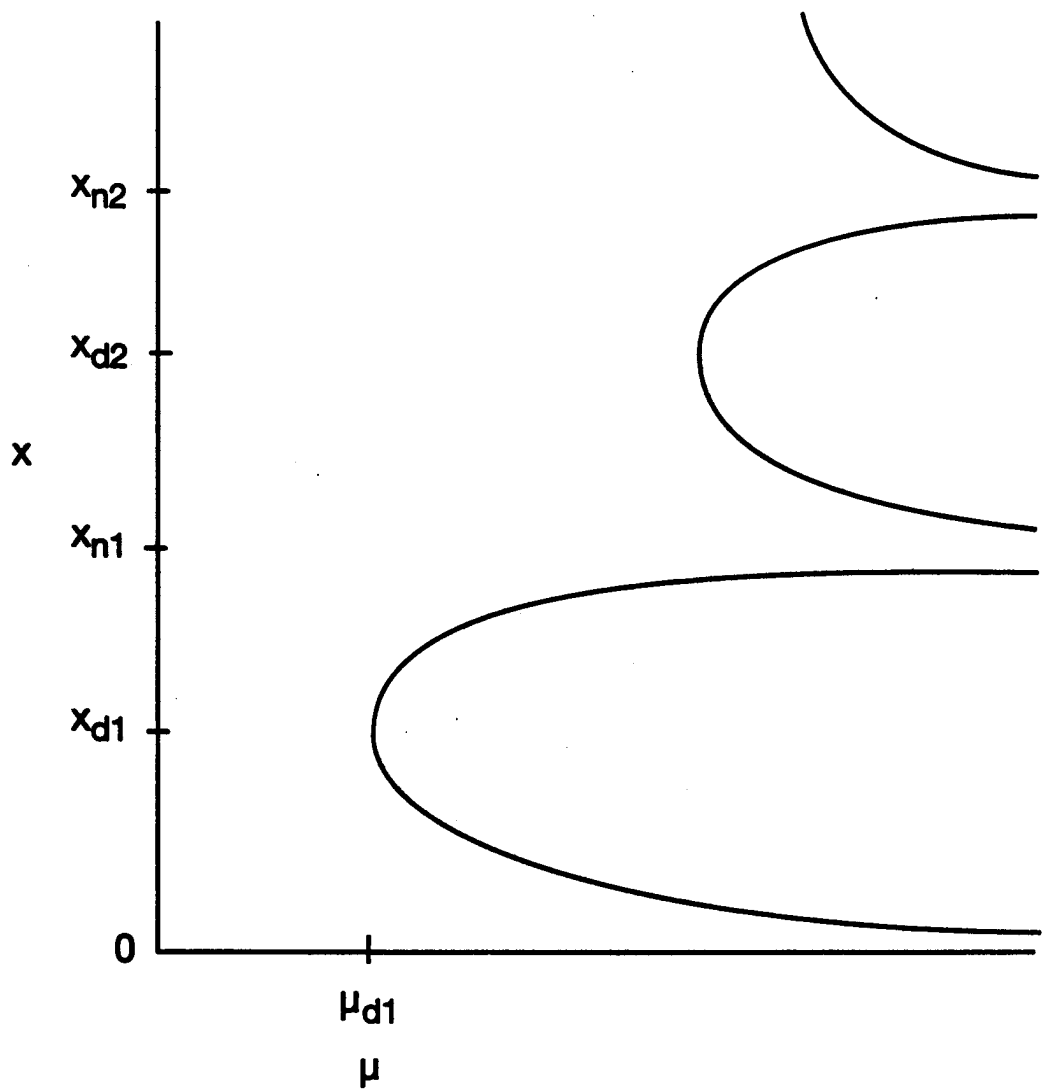
\item[Figure 2:] Schematic behaviour of $x(\mu)$ obtained from a
[2$\vert$2] Pad\'e-summation estimate of the $\beta$-function whose positive
numerator and denominator zeros alternate.  The denominator zeros are
denoted by $x_{d1}$ and $x_{d2}$, the numerator zeros are denoted by
$x_{n1}$ and $x_{n2}$, and the alternation of zeros is consistent
with the smallest positive zero being a zero of the denominator:  $0
< x_{d1} < x_{n1} < x_{d2} < x_{n2}$.  The value $\mu_{d1}$ is
defined such that $x(\mu_{d1}) = x_{d1}$;  values of $\mu < \mu_{d1}$
are outside the domain of $x(\mu)$.

\item[Figure 3a:]  Evolution of the running mass $m_b(\mu)$ obtained
from three-loop (3L) $\beta$- and $\gamma$-functions from the initial
condition $m_b(4.3 \; GeV) = 4.3 \; GeV$.

\item[Figure 3b:]  A comparison of the ratio of $m_b(\mu)$ obtained
from four-loop (4L) and [2$\vert$2] Pad\'e-summation $\beta$- and
$\gamma$-functions to $m_b(\mu)$ obtained from three-loop (3L)
$\beta$- and $\gamma$-functions, as shown in Fig. 3a, given the
initial condition $m_b (4.3 \; GeV) = 4.3 \; GeV$.

\item[Figure 3c:] A comparison of the evolution of $m_c(\mu)$
obtained from evolution of two-loop (2L), three-loop (3L) four-loop
(4L) and [2$\vert$2] Pad\'e-summation $\beta$- and
$\gamma$-functions.

\item[Figure 3d:] Masses of light (u, d, s) quarks from 2-loop,
3-loop, 4-loop, and [2$\vert$2] Pad\'e-summation $\beta$- and $\gamma$-functions.
Masses at $\mu$ = 1 GeV are normalized to unity.

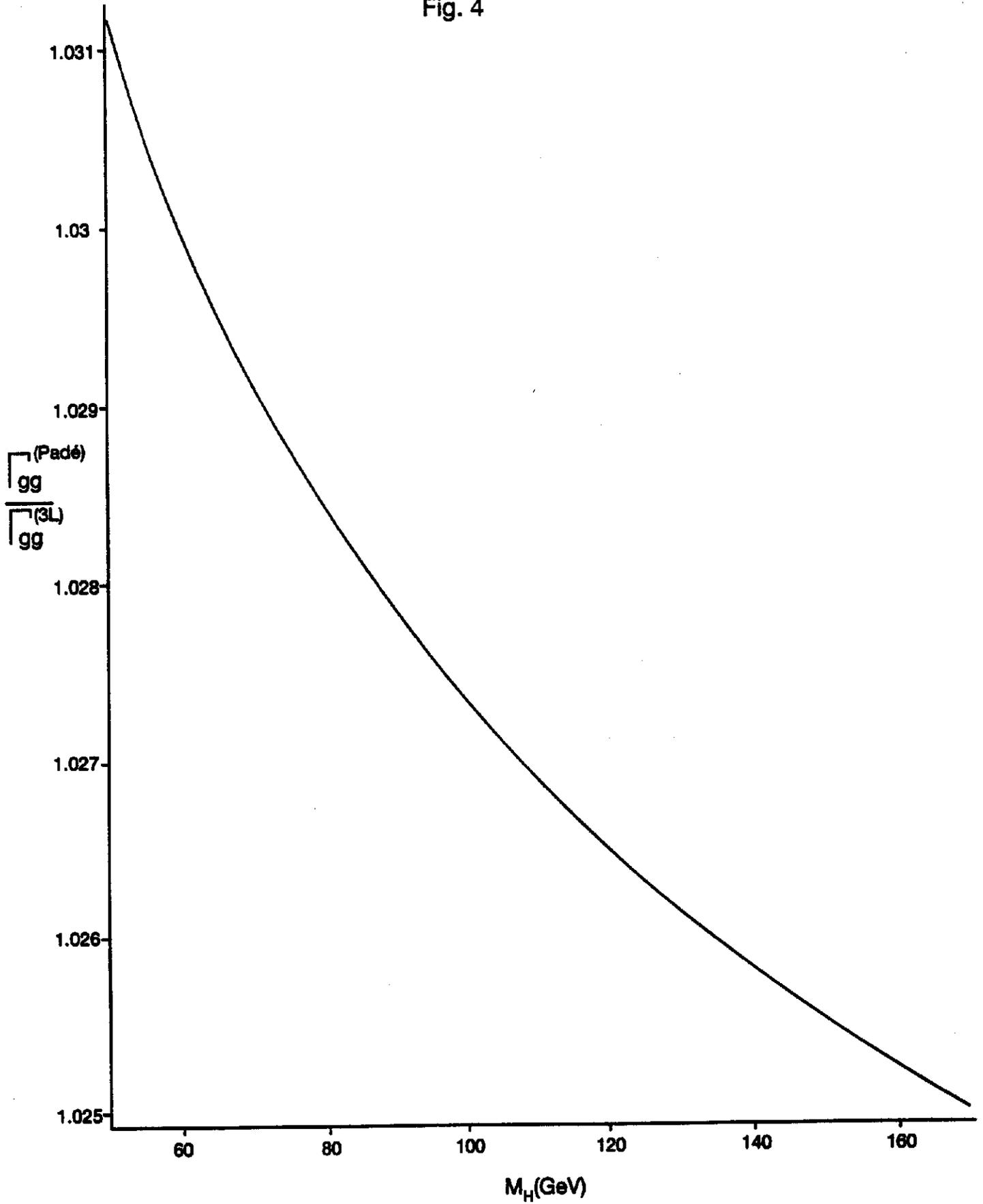
\item[Figure 4:]  A comparison of the ratio of the Higgs decay rate
into two gluons obtained via Pad\'e-improvement discussed in the text
to the three-loop (3L) expression for the same rate.

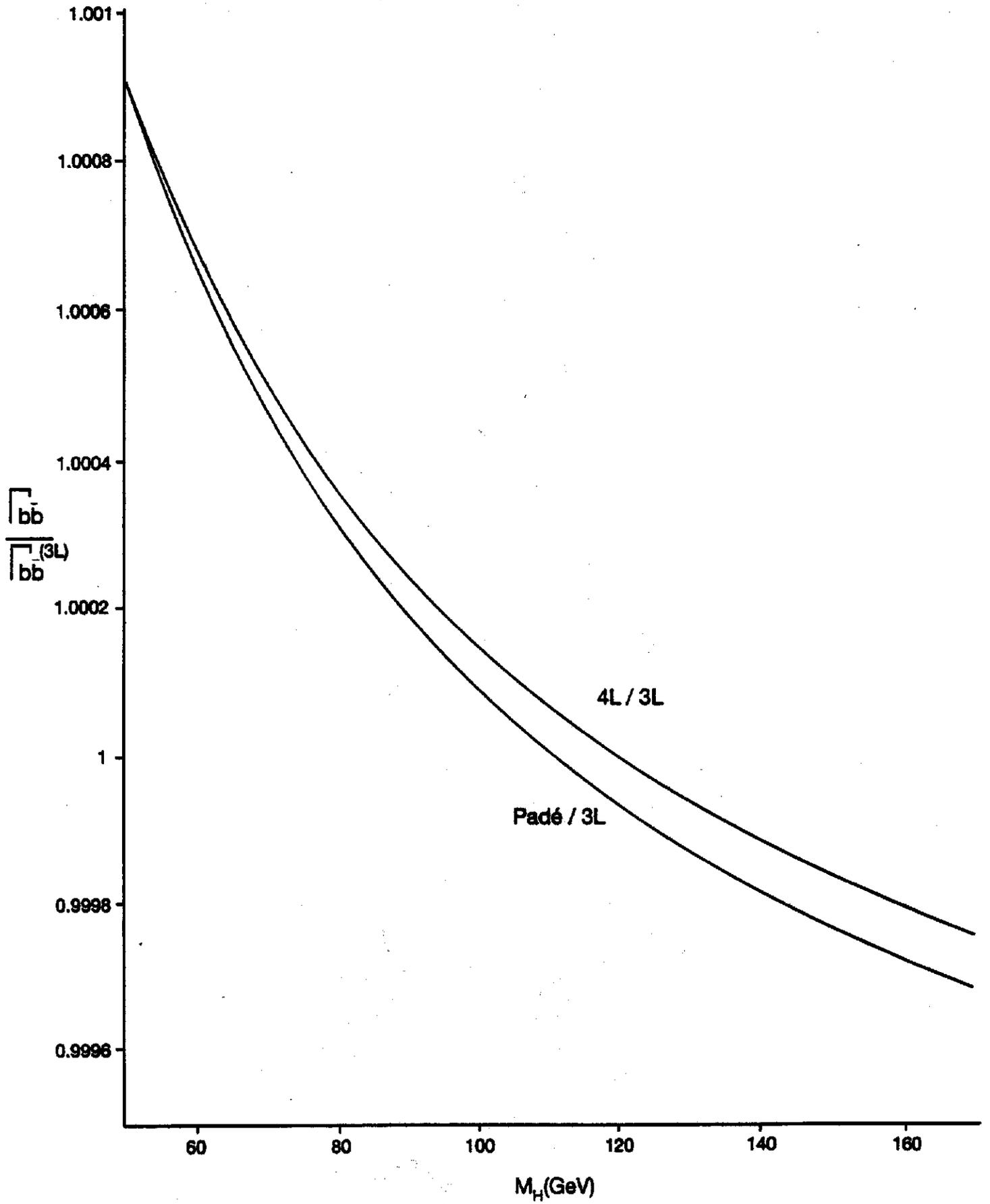
\item[Figure 5:]  A comparison of the ratio of the Higgs decay rate
into a $b \bar{b}$ pair obtained to four-loop (4L) order and through
subsequent Pad\'e-improvement (as described in the text) to the same
rate obtained to three-loop (3L) order.

\end{description}

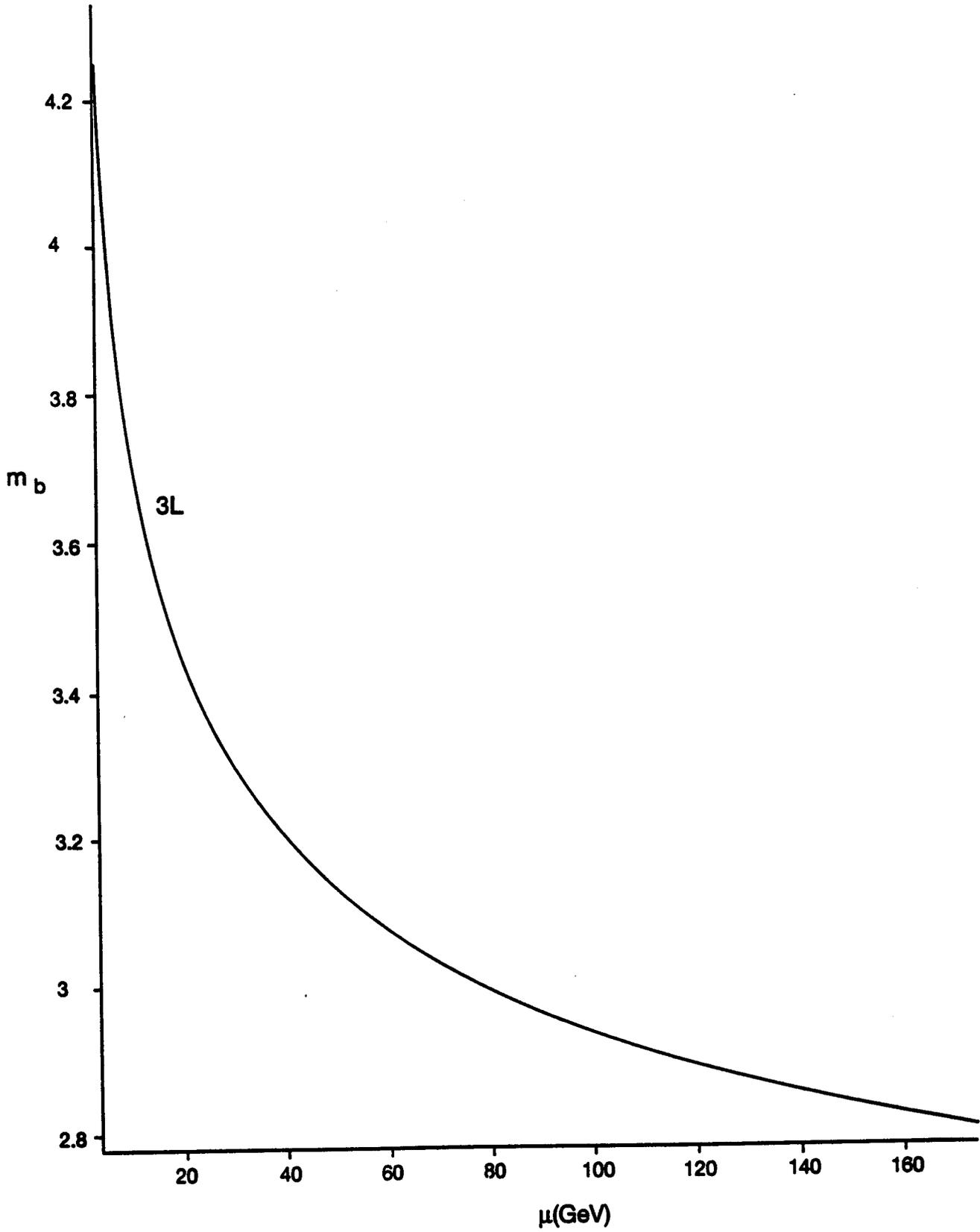
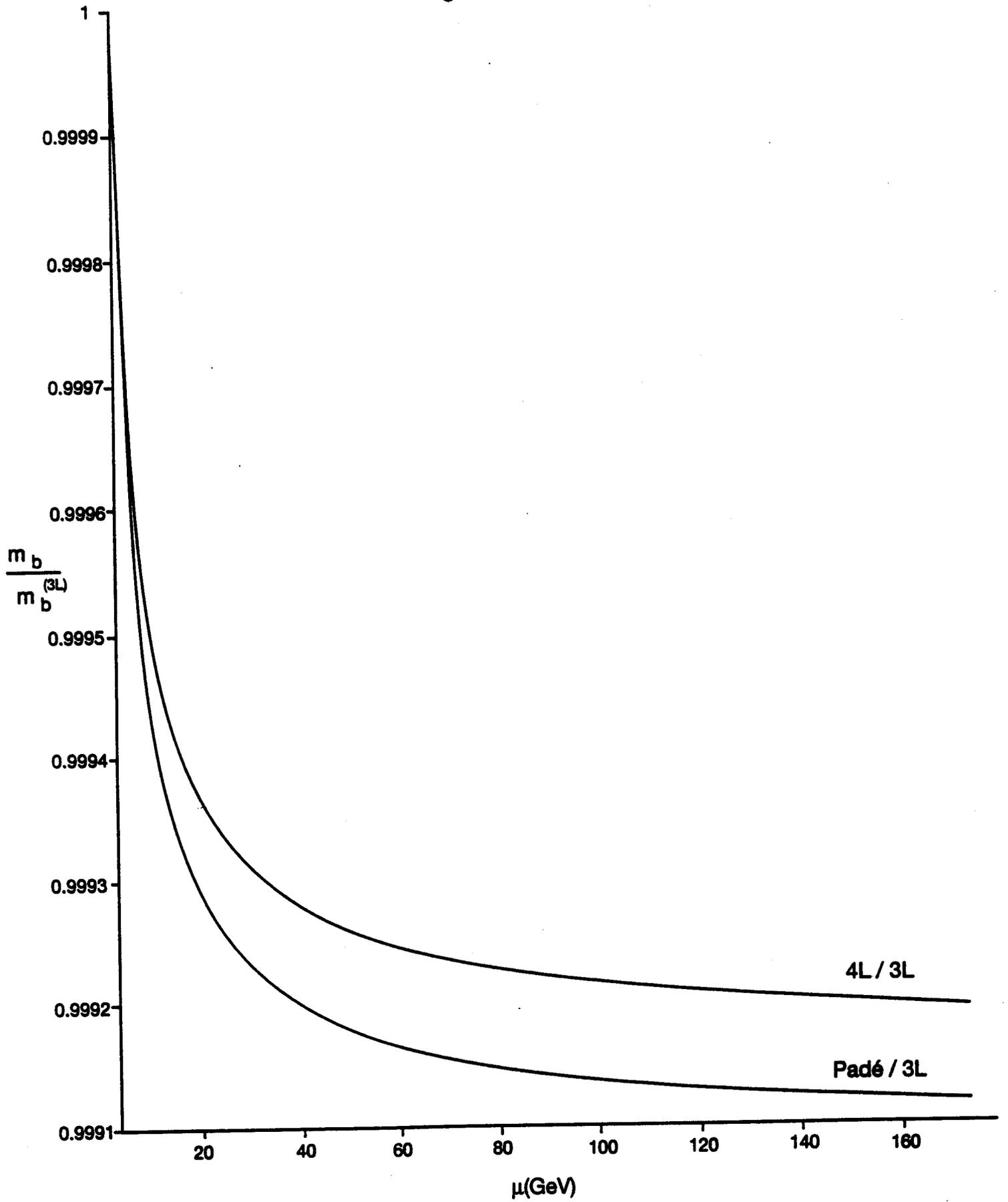
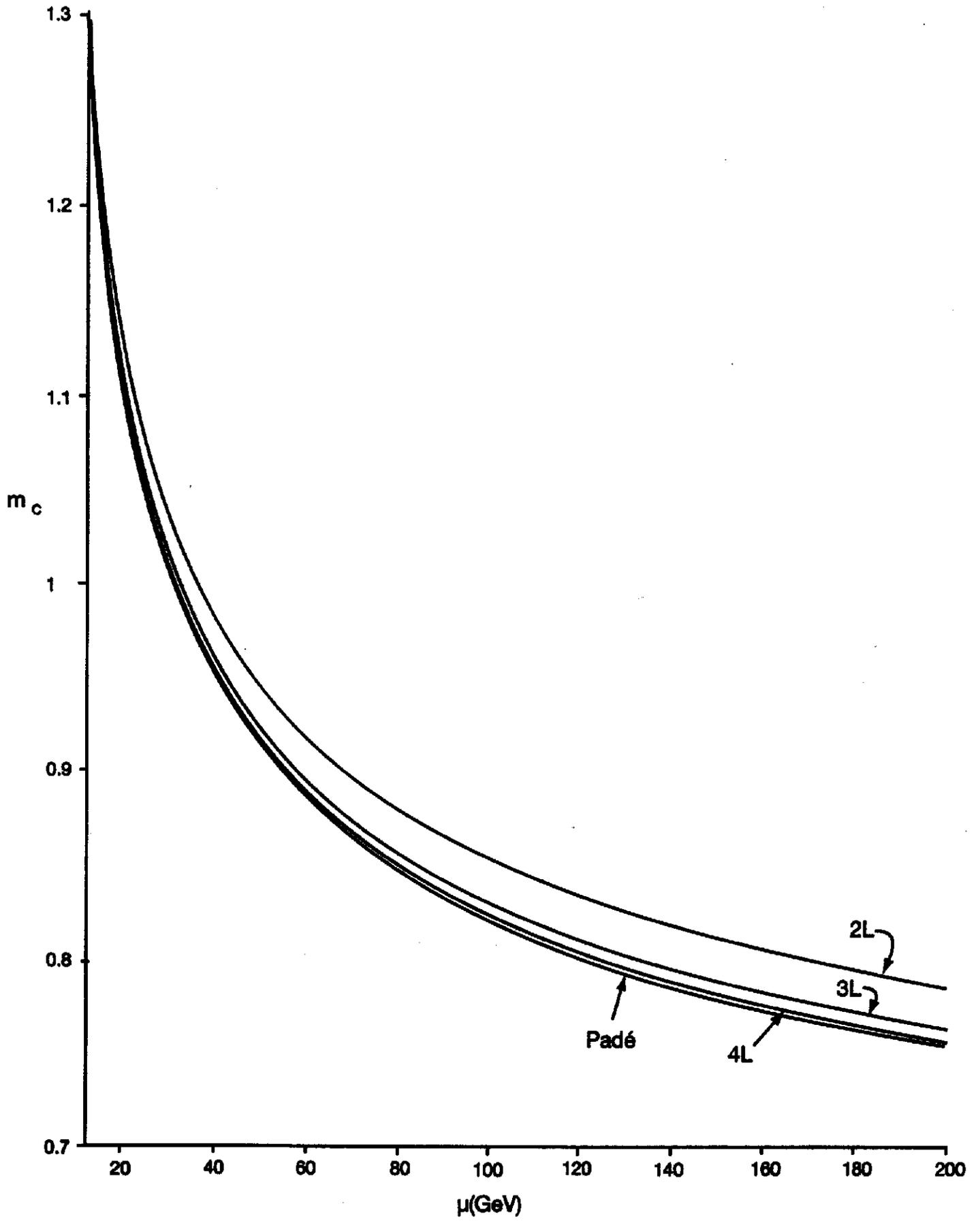
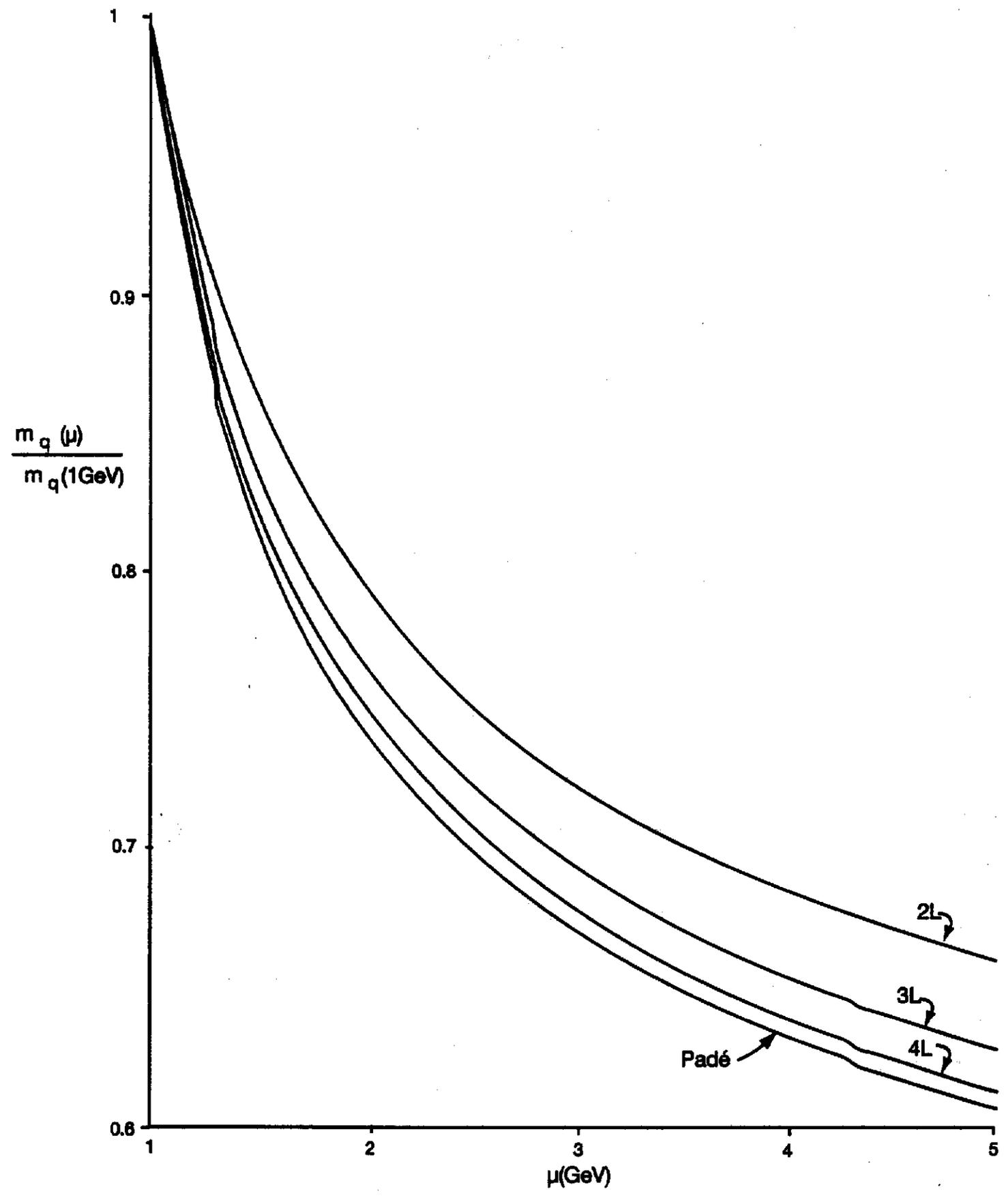
\end{document}